\documentclass[aps,preprint,superscriptaddress,longbibliography]{revtex4-1}
\usepackage{amsmath}
\usepackage{bm}
\usepackage{graphicx}
\usepackage{subfigure}
\usepackage{amsmath}
\usepackage{amssymb}
\usepackage{mathrsfs}
\usepackage{braket}
\begin{document}
\title{Solving  Hamiltonian Cycle Problem using Quantum $\mathbb{Z}_2$ Lattice Gauge Theory}
\author{Xiaopeng Cui}
\affiliation{Department of Physics \&  State Key Laboratory of Surface Physics, Fudan University, Shanghai  200433, China}
\author{Yu Shi}
\email{ yushi@fudan.edu.cn}
\affiliation{Department of Physics \&  State Key Laboratory of Surface Physics, Fudan University, Shanghai 200433, China}

\newcommand{\SCL}{0.6}
\newcommand{\COAO}{$O ( \frac{1}{g_c^2} \sqrt{  \frac{1}{\varepsilon}  N_e^{3/2}( N_v^3 +  \frac{N_e}{g_c} } ) ) $}
\begin{abstract}
   The Hamiltonian cycle (HC)  problem in graph theory is a well-known NP-complete problem. We present an approach in terms of $\mathbb{Z}_2$ lattice gauge theory (LGT) defined on the lattice with the graph  as its dual. When the coupling parameter $g$ is less than the critical value $g_c$, the ground state is a superposition of all configurations with closed strings of spins  in a same single-spin state, which can be   obtained by using an adiabatic quantum algorithm with time complexity $O(\frac{1}{g_c^2} \sqrt{  \frac{1}{\varepsilon}  N_e^{3/2}(N_v^3 +  \frac{N_e}{g_c}}))$, where  $N_v$ and $N_e$ are the numbers of vertices and  edges of the graph respectively. A subsequent search for a HC among those closed-strings solves the  HC problem.    For some random samples of small graphs, we demonstrate that the dependence of  the average value of $g_c$ on $\sqrt{N_{hc}}$,  $N_{hc}$ being the number of HCs,  and that of the average  value of $\frac{1}{g_c}$ on  $N_e$ are both linear. It is thus suggested that for some graphs, the HC problem may  be solved in polynomial time. A possible quantum  algorithm using $g_c$ to infer $N_{hc}$  is also discussed.


\end{abstract}

\maketitle

In computational complexity, an important yet unproven conjecture for  classical computing  is P $\neq$ NP, where P represents the class of problems that can be solved in times  polynomial in the sizes of the inputs, while  NP is the shorthand for nondeterministic polynomial,  representing the class of problems the  solutions  of  which can be verified in polynomial times.  P is a subset of NP. The factoring problem, which is to find the prime factors of an odd number, is in class  NP, but is conjectured not to be in class P.

In quantum computing,  the factoring problem is in  class P, as it  can be solved by using Shor's  algorithm  in polynomial times~\cite{Shor1994}, suggesting that quantum computing is more powerful than classical computing. However, it cannot be concluded  that  P=NP  in quantum computing,  because factoring, as a specific NP problem,  is not a  NP-complete  (NPC) problem, to which  any NP problem  can be reduced to in polynomial time. If any  NPC problem   can be solved in  polynomial time, so do all the  NP problems and consequently P=NP.

Therefore, if  there is a quantum algorithm of polynomial time  for any NPC problem,  it can be concluded that P=NP in quantum computing. It is thus important to find quantum speedup  of any NPC problem.

A well-known  NPC problem is the HC problem~\cite{van}. On an undirected graph, which is made up of a number of  vertices connected by a number of  edges without directions~\cite{Garey}, a HC  is a cycle  that visits each vertex exactly once. HC problem  is to determine  whether such a HC exists. So far the best classical algorithms include the so-called dynamic programming algorithm of time complexity $O({N_v}^2 2^{N_v})$~\cite{Bellman1962} and a Monte Carlo algorithm of time complexity $O(1.657^{N_v})$~\cite{hcp2010}, both of which are exponential in the number of vertices $N_v$. Up to now there is  neither quantum algorithm with  P time complexity~\cite{hcp_quantum2019}.

In this paper,  we propose a new approach to HC  problem and thus the problem of P versus NP. We   transform the HC problem to a problem in  quantum  $\mathbb{Z}_2$ LGT~\cite{Kogut1979,sachdev}. Gauge theory is the framework describing the fundamental interactions in   particle physics, as well as strong correlations in condensed matter physics.  LGT is an approach to gauge theories by discretizing the  spacetime or space to  a lattice.  $\mathbb{Z}_2$ LGT is the simplest LGT. Recently quantum simulation has become a new approach to LGT, including  $\mathbb{Z}_2$ LGT~\cite{Lloyd1996,RN669, qz2_cui,Lamm2019}.

\begin{figure}[htb]
   \centering
   \includegraphics[scale=\SCL]{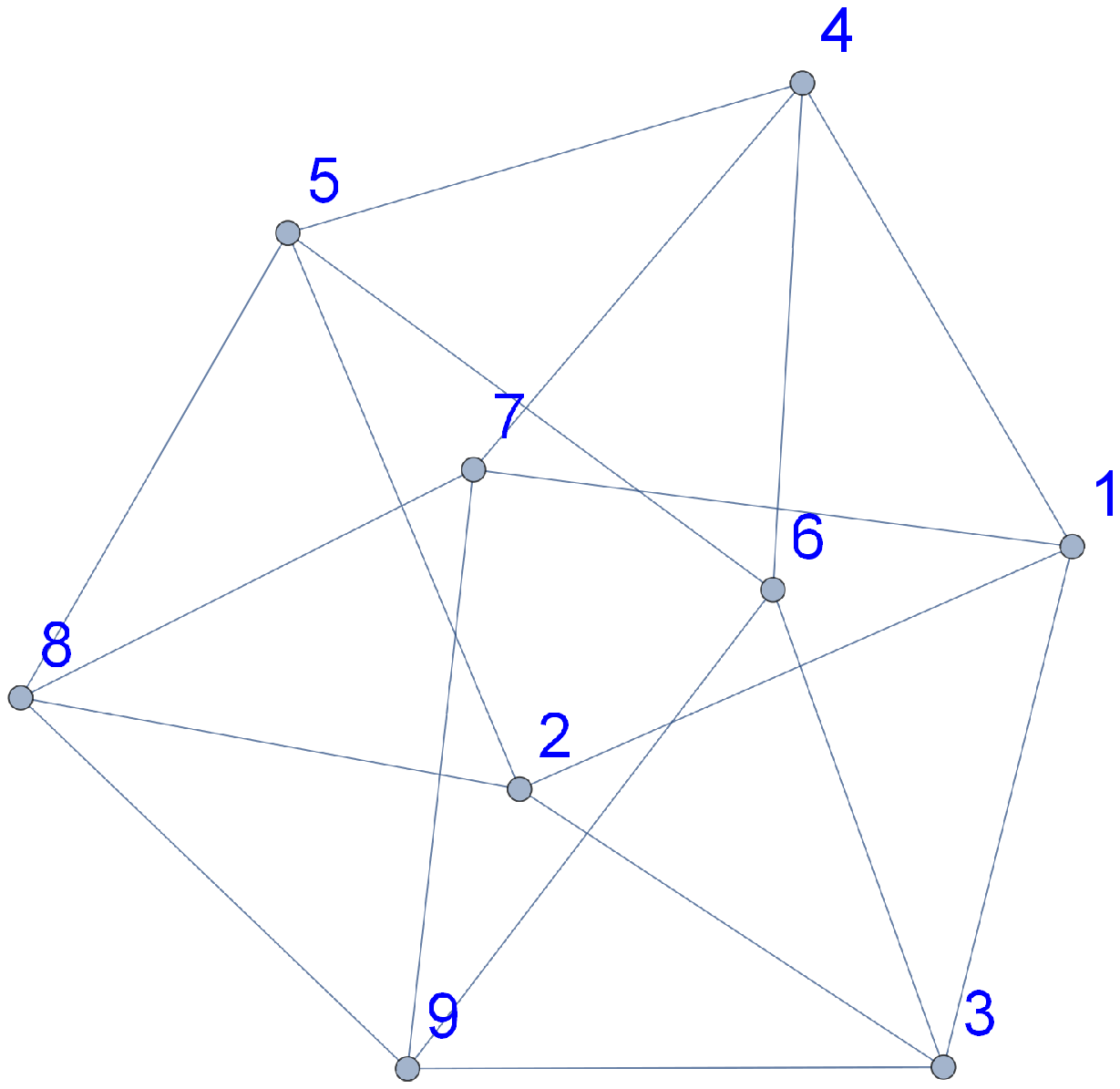}
   \subfigure[]{}
   \includegraphics[scale=\SCL]{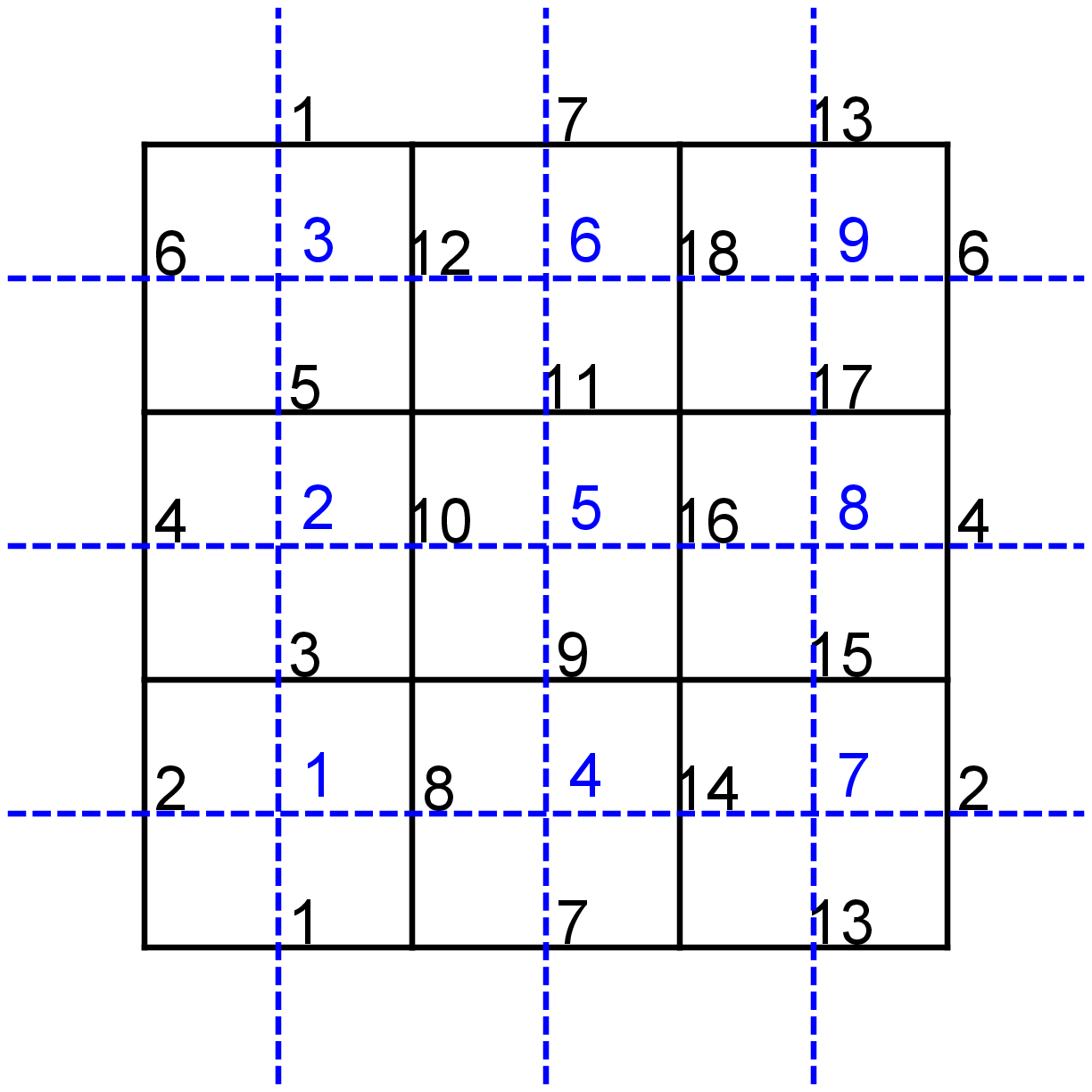}
   \subfigure[]{}
    \caption{Correspondence between a
    graph and the torus lattice.   (a)  A   graph corresponding to the $3\times3$ tours lattice shown in (b).   (b)  $3\times3$ torus lattice, where the black solid lines are links of the lattice,  the blue dotted lines correspond to the edges of the graph in (a), the numbers in black denote the edges of the lattice, and the numbers in blue correspond to those for the vertices in  the graph in (a). }
   \label{fig_torus_graph}
\end{figure}

We regard a graph as the dual  of a lattice, on which the $\mathbb{Z}_2$ LGT is defined.  Thus each vertex of the graph maps to a  plaquette  in the lattice,   each edge in  the graph crosses a link in the lattice, and two adjacent vertices in the graph correspond to two adjacent plaquettes in the lattice. The length is not defined for  each edge in the graph or each link in the lattice, therefore  the lattice can be deformed when needed. Thus an undirected unweighted graph of $N_v$ vertices becomes the dual lattice  of a  lattice with  $N_v$ palquettes.  Depicted  in Fig.~\ref{fig_torus_graph} is an example, in which the graph becomes   the dual lattice of a $3\times 3$ square lattice,  with periodic boundary condition, i.e. a topologically torus lattice.

\begin{figure}[htb]
   \centering
      \subfigure[]{}
      \includegraphics[scale=\SCL]{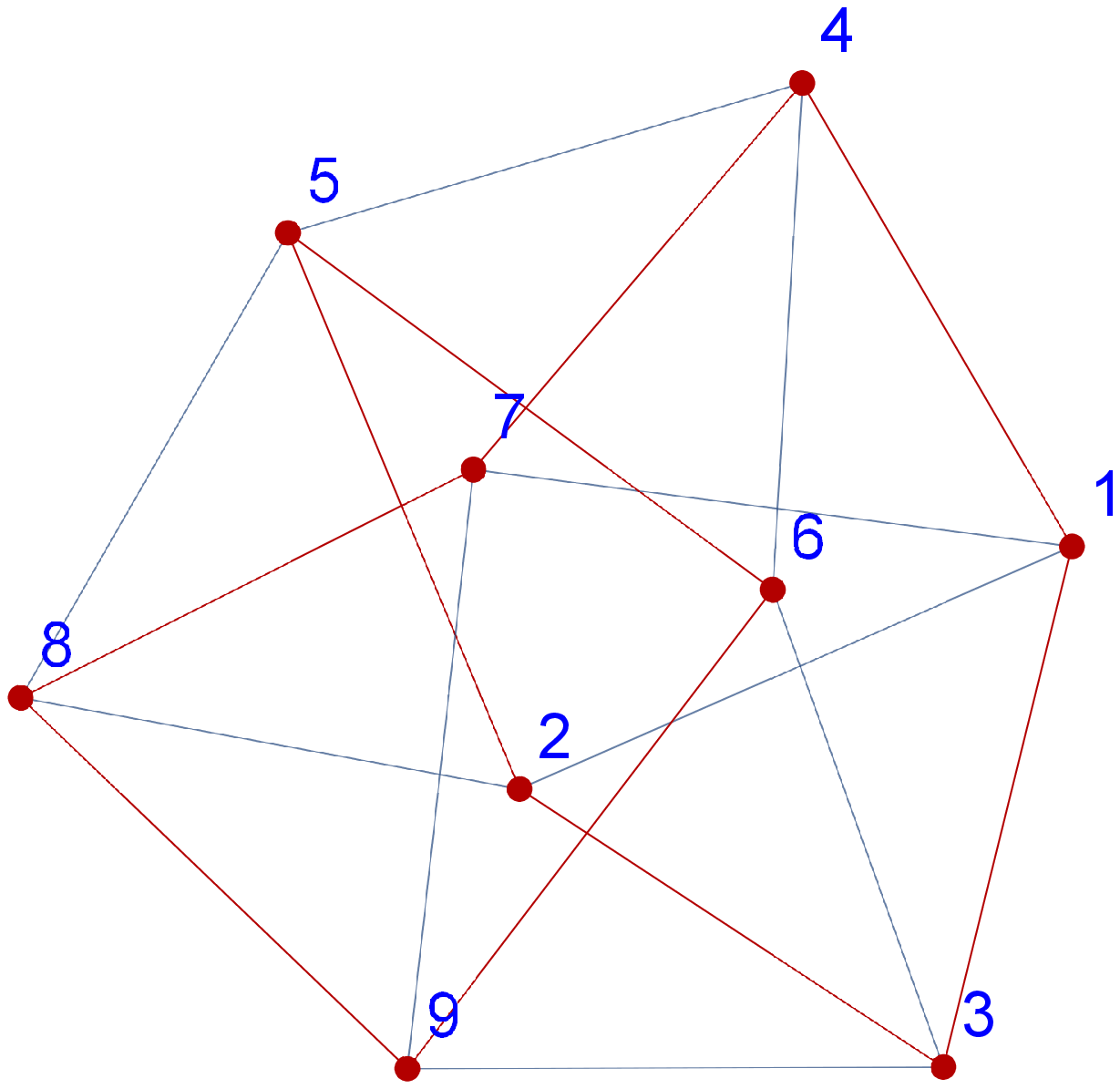}
      \subfigure[]{}
      \includegraphics[scale=\SCL]{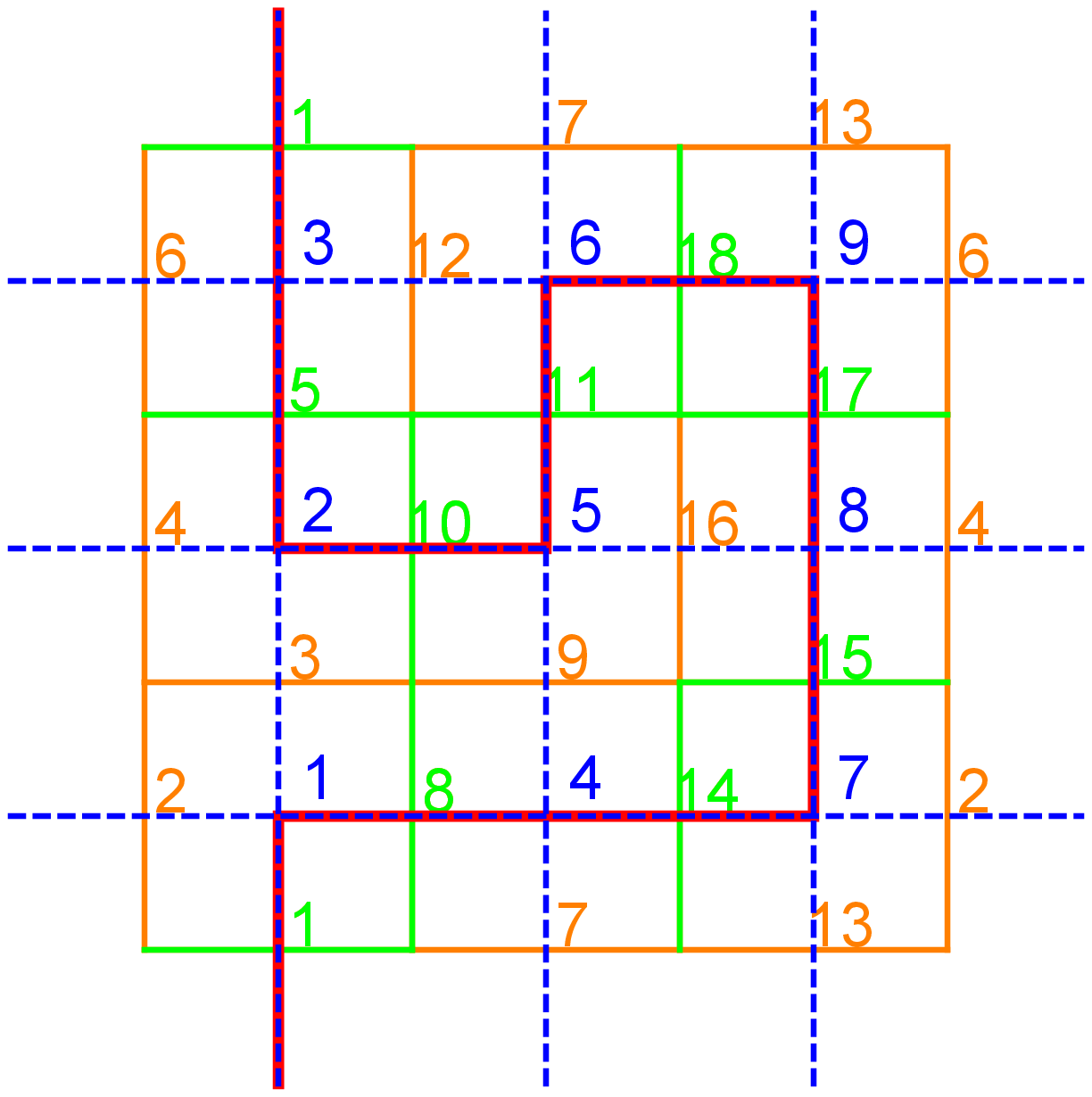}

      \subfigure[]{}
      \includegraphics[scale=\SCL]{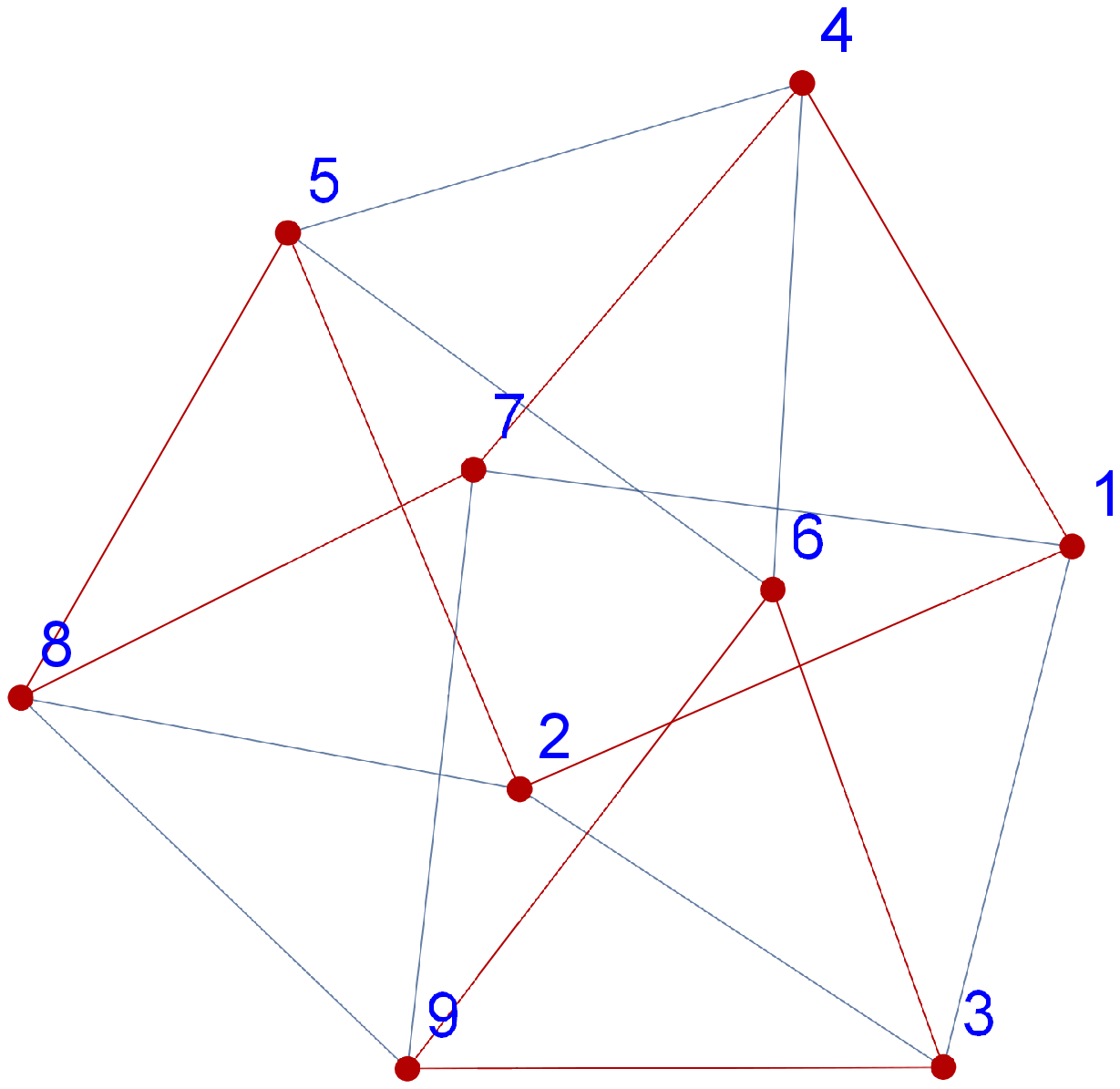}
      \subfigure[]{}
      \includegraphics[scale=\SCL]{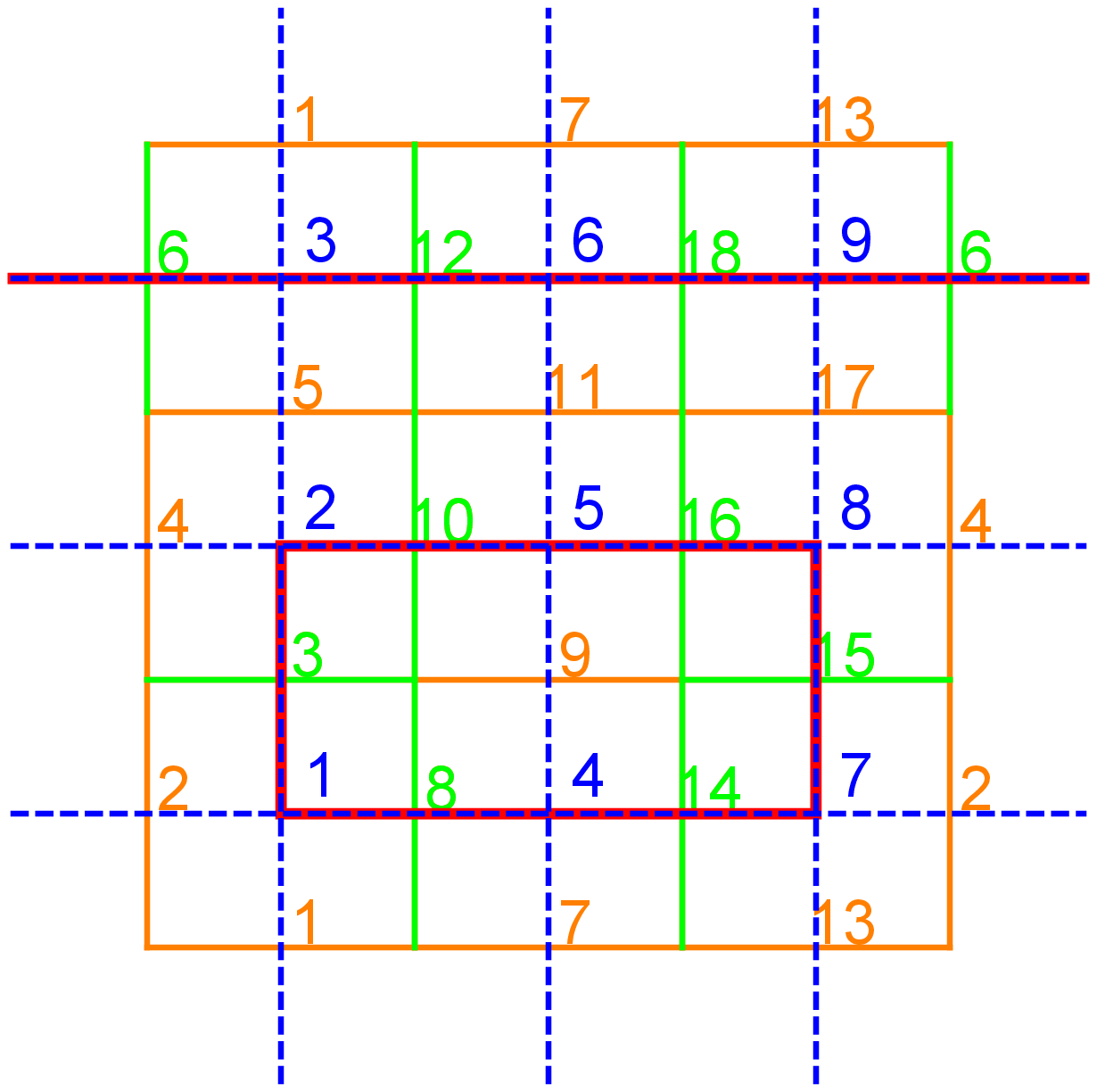}

      \caption{Identification of the cycles on a certain graph  as the closed strings on the dual lattice  of a square lattice.  (a) A  HC of the graph, depicted as a red path in (a) becomes  a closed string  passing through all the plaquettes on the lattice,   as depicted in red in (b). (b) A  configuration of closed strings at the ground state of quantum $\mathbb{Z}_2$ LGT at $g=0$. There is only one closed string, depicted as a red path.  The orange edge is in the $\ket{0}$ state, and the green edge is in the $\ket{1}$ state.   (c) Two cycles, depicted in  red,   in  the graph.    (d) A closed string configuration in the quantum $\mathbb{Z}_2$ LGT ground state at g = 0, with two closed strings, depicted as  red paths, corresponding to the two cycles in the graph. The orange edge is in the $\ket{0}$ state, and the green edge is in the $\ket{1}$ state.  }
      \label{fig_torus_graph_cycle}
   \end{figure}

In this way, HC in a graph becomes HC in the dual lattice of a lattice.    By connecting the last with the first visited vertices in the graph,  an HC in the dual lattice becomes a closed string or loop  passing through all the plaquettes of the lattice  with periodic  boundary condition, as shown in Fig.~\ref{fig_torus_graph_cycle}. Hence HC problem becomes the search for a closed string that visits each vertex of the dual lattice one. This is much easier, since the   set of all the closed strings  is much smaller than the set of all paths.

Furthermore, the search for a HC among the closed strings can be significantly accelerated by the topological quantum phase transition (TQPT) in the quantum $\mathbb{Z}_2$ LGT.  At the critical value $g_c$ of the coupling parameter $g$, a TQPT occurs separating  the confined phase for $g > g_c$ and the  deconfined phase for $g <g_c$, in  which there is proliferation of the closed strings of down spins, called closed-string condensation, as the ground state becomes a closed-string condensate, which is a quantum  superposition of all  configurations of closed strings of down spins (in the convention described below)~\cite{Levin2005,Wen2005}.  $g_c$ depends on the properties of the  graph, including $N_{hc}$, hence can be used to reveal  $N_{hc}$. Therefore, reaching TQPT in a quantum adiabatic simulation becomes the key step in the present approach to HC problem. The dependence of  $g_c$ on $N_{hc}$ suggests low time complexity, and to a new approach to the problem of P versus NP. This method applies to any graph,  which can always be regarded as a dual lattice of a lattice, on which the $\mathbb{Z}_2$ LGT can always be defined.


{\em Quantum  $\mathbb{Z}_2$ LGT and closed strings.----}The Hamiltonian of the quantum $\mathbb{Z}_2$ LGT~\cite{qz2_cui, sachdev, Wegner1971} is
\begin{equation}
    H  = Z + gX
   \end{equation}
   with
\begin{equation}
   X= -\sum_l { \sigma_l^x },\, Z = \sum_{\square} { Z_\square} ,\, Z_\square= -\prod_{l \in \square}{  \sigma_l^z }
\end{equation}
where  the spins (qubits) occupy the links $l$'s,  $\square$ represents the elementary plaquette of the lattice. Single-spin states $|0\rangle$ and $|1\rangle$ correspond to  $\sigma_l^z= 1$ (up) and  $\sigma_l^z=-1$ (down), respectively. The  ground state    at $g=0$  is
\begin{equation}
\ket{\psi_0} = \sum_{i=1}^{S_0} { \ket{\phi_i} }, \label{p0}
\end{equation}
where each  ${ \ket{\phi_i} }$ is  with   $Z_\square = -1$ for any $\square$ and thus  $Z=-9$. In the $M\times N$ torus lattice, there are  $N_v\equiv MN$ plaquettes, $N_e\equiv 2N_v$  links. Hence the number of configurations with  $Z=-9$ is $S_0 \equiv  2^{N_e}/(2^{N_v-1})$,  where $2^{N_e}$ is  the number of all possible configurations of  spins, while $1/2^{N_v-1}$ is the probability that $Z_\square = -1$ for each $\square$.

On the dual lattice, one can regard a spin  with $\sigma_l^z=-1$ as occupied by a string while  a link with $\sigma_l^z= 1$ as unoccupied. Occupied links can be  connected to form a longer string, which may be open or closed, then   $\ket{\phi_i}$  can  be regarded as  a configuration of closed strings. When $g<g_c$, the ground state can be represented as a closed-string condensate, which is a superposition of $S_0$ configurations of closed strings. Especially, when $g=0$, the superposition is of equal amplitude~\cite{freedman,Levin2005,fradkinbook},   Among the $S_0$ configurations of closed strings, two examples are as shown in the Fig.~\ref{fig_torus_graph_cycle}, where in (a), a single closed string passes through all $N_v$ plaquettes of the lattice, and is thus a HC, while in (b), there are two closed strings, each  passing through a part of the $N_v$ plaquettes, so there is no HC.

Therefore, in order to solve HC problem, one only needs to search for HCs  in the  $S_0$ configurations of closed strings, rather than  $N_v!$ configurations in  the original configuration space~\cite{Garey}. This greatly reduces complexity.

{\em Quantum Algorithm for HC problem.----}The complexity of this quantum algorithm comes from the  two consecutive processes. The first is  to obtain  the closed-string condensate, and the second is  to  search  for HCs in it.

The initial state is   the ground state at $g=+\infty$, which is the equal superposition of all product states of $|0\rangle$ and $|1\rangle$.  With $g$ decreased  adiabatically towards $g \leq g_c$, the  closed-string condensate  is obtained. This process is equivalent to adiabatically evolve $H_\lambda= \lambda Z + X$,  with $\lambda=\frac{1}{g}$, from  $\lambda = 0$ towards $\lambda_c =\frac{1}{g_c}$.

The adiabatic condition  implies that the time scale is $t=O(\sqrt{N_e})$ for  a 2-dimensional lattice \cite{RN693}. Hence the time scale is $t=O(\sqrt{N_e})\lambda_c$ for an lattice of an unknown $\lambda_c$.

For the adiabatic evolution, we use the symmetric Trotter decomposition to decompose the unitary evolution, which is more efficient than the original Trotter decomposition \cite{Trotter1959, Childs_2019}.  The error for each step in  the adiabatic simulation is
 $\varepsilon_s(t,n,g)  =| e^{-i(A+B)t} - (e^{-iA\frac{t}{2n}}e^{-iB\frac{t}{n}} e^{-iA\frac{t}{2n}})^n  |                                         =| e^{-i(A+B)t} - e^{-iA\frac{t}{2n}}  (e^{-iB\frac{t}{n}}e^{-iA\frac{t}{n}})^{n-1}e^{-iB\frac{t}{n}} e^{-iA\frac{t}{2n}}|
                           \leqslant ( \frac{1}{12} g^2  N_v n_l^2 + \frac{1}{24} g N_e*n_p^2) \frac{t^3}{n^2}$,                                            
where $A=Z$ and $B=gX$ are the two terms in the Hamiltonian,  $t$ is  time of the step, $n$ is number of symmetric Trotter substeps, $n_l$ is the number of links surrounding each plaquette,  $n_p$ is the number of vertices connected by each edge or is the number of plaquettes sharing each link~\cite{qz2_cui}.  Then
the cumulative error from the start to the phase transition point $\lambda_c$ can be obtained as
\begin{equation}
   \varepsilon = O( \frac{1}{N_{ss}^2}  N_e^{3/2}(N_v^3+N_e\lambda_c) \lambda_c^4  ),
\end{equation}
where $N_{ss}$ is the total number of symmetric Trotter substeps. Therefore,  the time complexity required in the  adiabatic quantum simulation of the quantum $\mathbb{Z}_2$ LGT   is
\begin{equation}
   O_1= O( \sqrt{\frac{1}{\varepsilon}  N_e^{3/2} (N_v^3+N_e\lambda_c) \lambda_c^4}  )= O(\frac{1}{g_c^2} \sqrt{  \frac{1}{\varepsilon}  N_e^{3/2}( N_v^3 +  \frac{N_e}{g_c} } ) ),
\end{equation}
where  $\varepsilon$ is reinterpreted as the precision required for the quantum simulation. In a connected undirected graph, $N_v \leq N_e\leq N_v(N_v-1)/2$,  $g_c$, $\lambda_c$ are fixed quantities for  a given graph.

\begin{figure}[htb]
   \centering
   \subfigure[]{}
   \includegraphics[width=0.45 \textwidth]{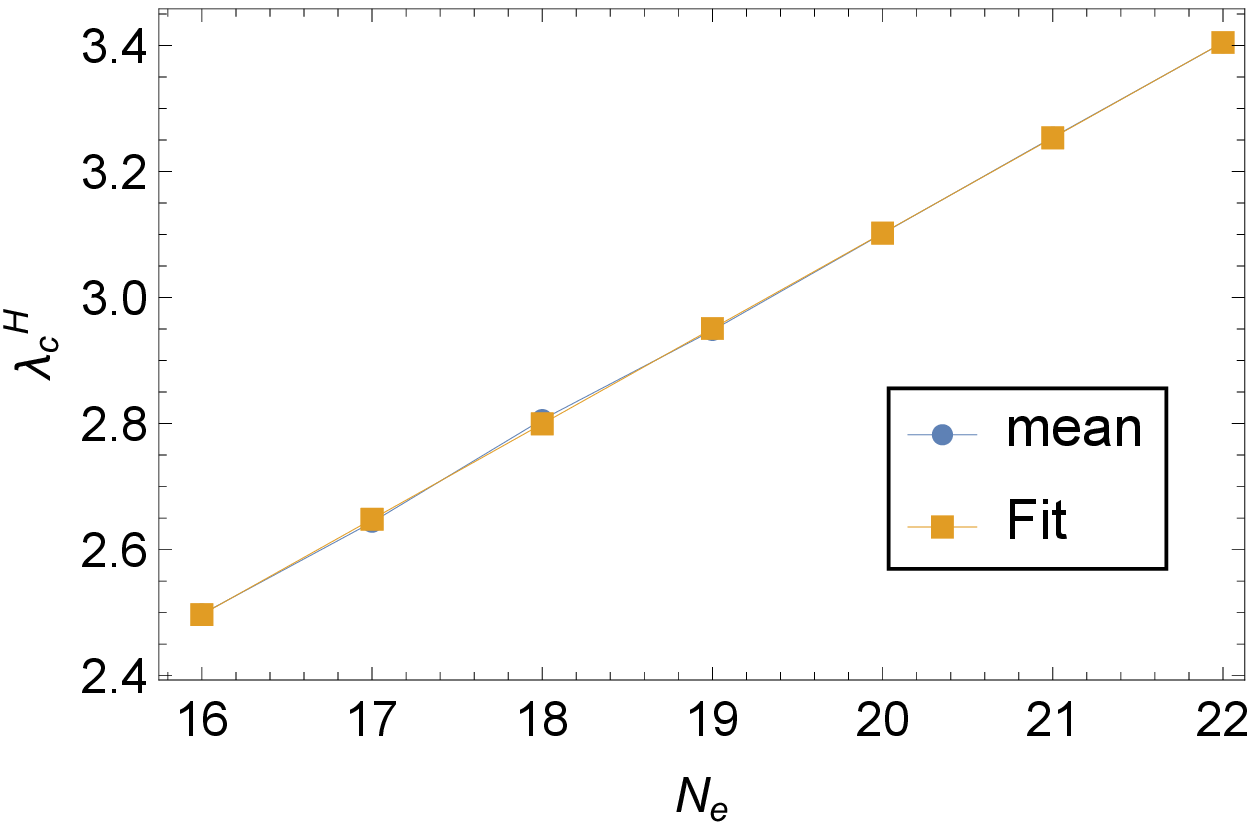}
   \subfigure[]{}
   \includegraphics[width=0.45 \textwidth]{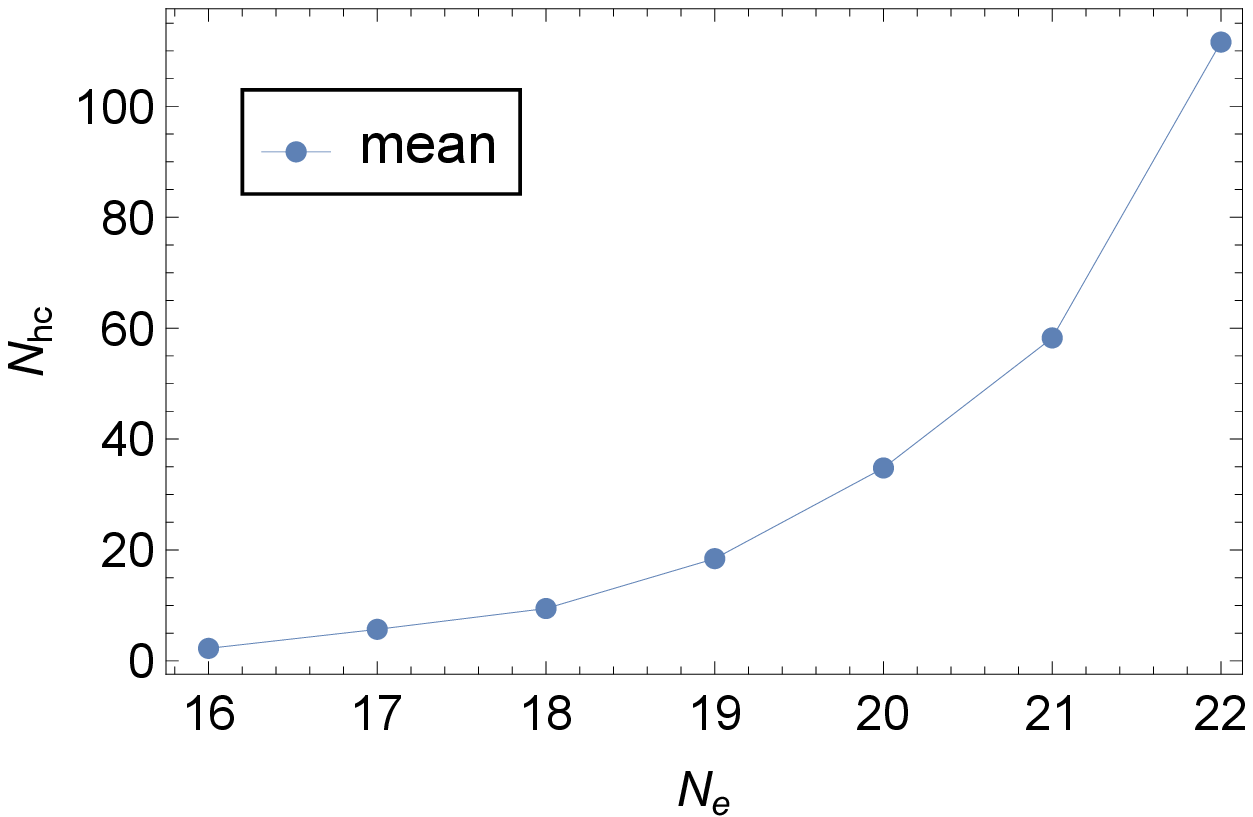}
   \caption{  (a) $\lambda_c^H$  as a function of  $N_e$, fitted by $\lambda_c^H = 0.1513*N_e + 0.07536 $.   $N_e$ varies from  $16$ to $22$.    (b) The dependence of the average  value of $N_{hc}$ on  $N_e$, which  varies from  $16$ to $22$.}
   \label{fig_hpc_lam_fit}
\end{figure}

Subsequently, a quantum search algorithm searches for HCs in the closed-string condensate, which is is a   superposition of $S_0 =O( 2^{N_e}/2^{N_v})$ components, including those with  HCs.  The time complexity  is $O_2\sim \frac{S_0}{N_{hc}}$, since   the probability of  obtaining a state with a HC  in a  measurement is $\frac{N_{hc}}{S_0}$.  The relationship between  $N_ {hc}$ and $N_e$  of some sample graphs is shown in figure \ref{fig_hpc_lam_fit} (b). It is   possible to develop a better quantum search algorithm finding HCs in the closed-string condensates.
The  time complexity  for the whole process is   $O=O_1\cdot O_2$. 

{\em Quantum Algorithm of finding $g_c$.----}$g_c$ depends on and thus may reveal some  properties of the graph, for example,   $N_{hc}$. After reaching the TQPT, a measurement of $g_c$   contributes  additional  time complexity $O_M$. According to the phase estimation method \cite{qz2_cui}, $O_M$ is of the same order of magnitude as $O_1$.  Thus with adiabatic algorithm followed by measurement, we have a quantum algorithm of finding $g_c$, with time complexity $O'=O_1+O_M \sim O_1$.

{\em Relations between $g_c$ and the graph properties.----}Because the complexity $O_1$ depends on $\lambda_c \equiv \frac{1}{g_c}$,  we need to investigate the relations between $g_c$ and the graph properties, e.g. $N_{hc}$ and  $N_e$.

We work on this problem using the  classical demonstration of the quantum adiabatic simulation, in terms of  the QuEST simulator~\cite{quest}. The hardware we use  is  the  Nvidia GPU Tesla V100-SXM2-32GB~\cite{qz2_cui}.

We first randomly generate  four graphs with $N_v=9$ and $N_e=18$, as  shown in Fig. \ref{fig_hpc_list_graph}(a-d). The fourth corresponds to the $3\times 3$ torus lattice.  $N_{hc}$ of the four graphs are 0, 10, 30, 48,  respectively. For such small graphs,  we use measurement method,  as described in \cite{qz2_cui}, to obtain the  ground state $\ket{\psi_0}$ at $g=0$, as given in \eqref{p0}. Note that this method is not feasible for larger graphs, as in our discussion on the quantum algorithm for HC problems.

\newcommand{\SCLL}{0.5}

\begin{figure}[htb]
   \centering
   \subfigure[]{}
   \includegraphics[scale=\SCLL]{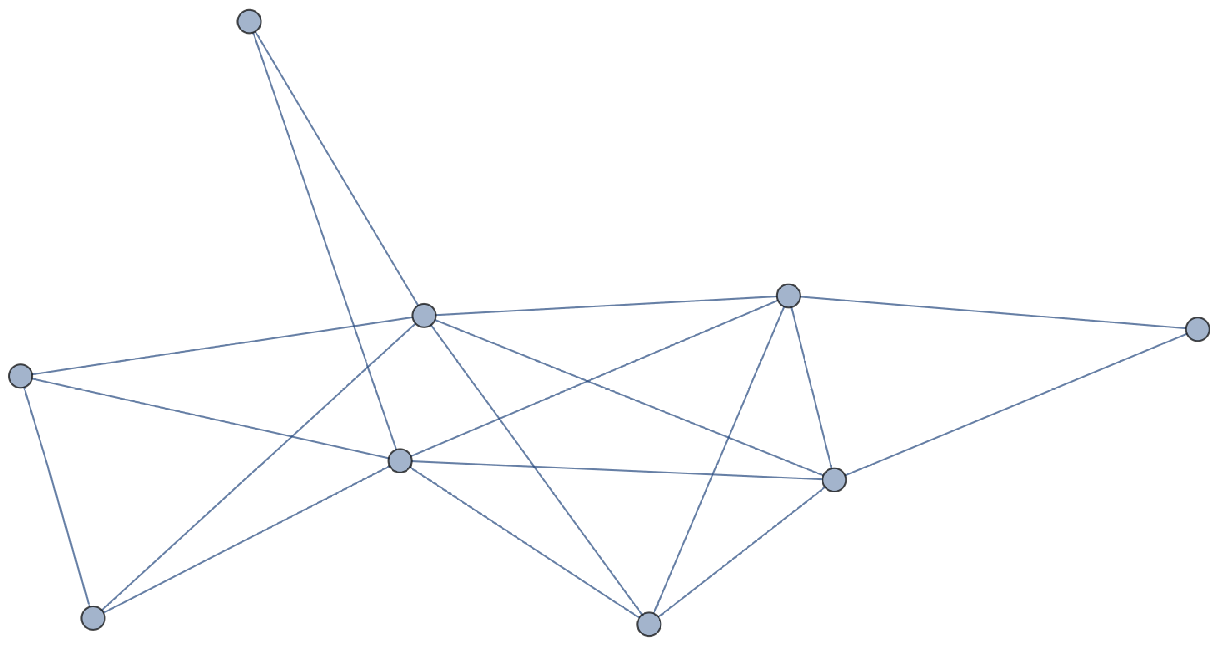}
   \subfigure[]{}
   \includegraphics[scale=\SCLL]{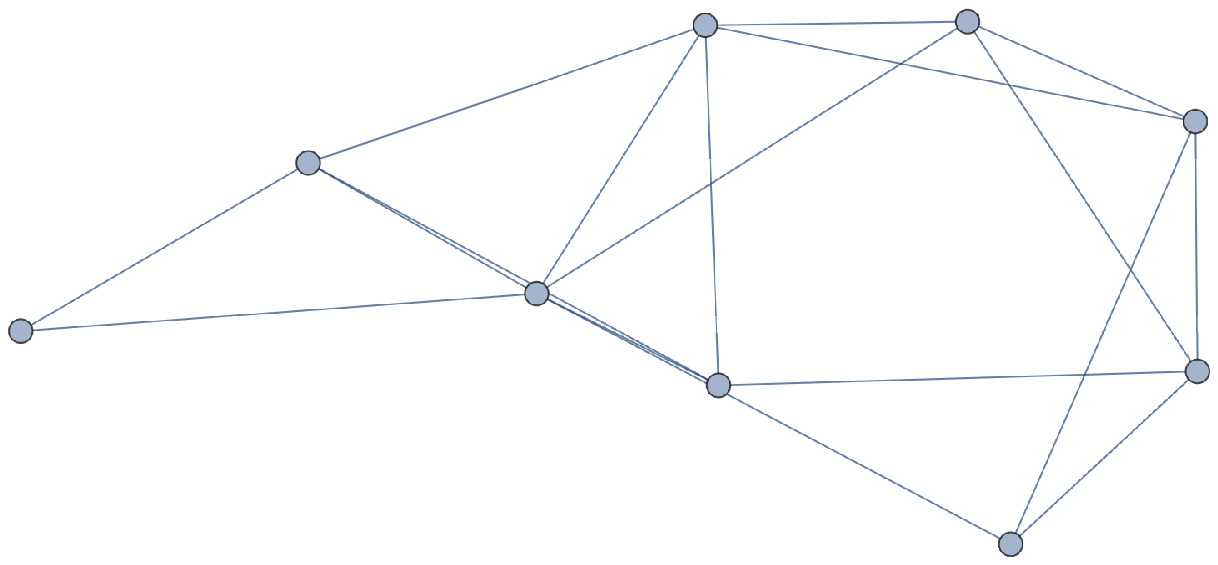}

   \subfigure[]{}
   \includegraphics[scale=\SCLL]{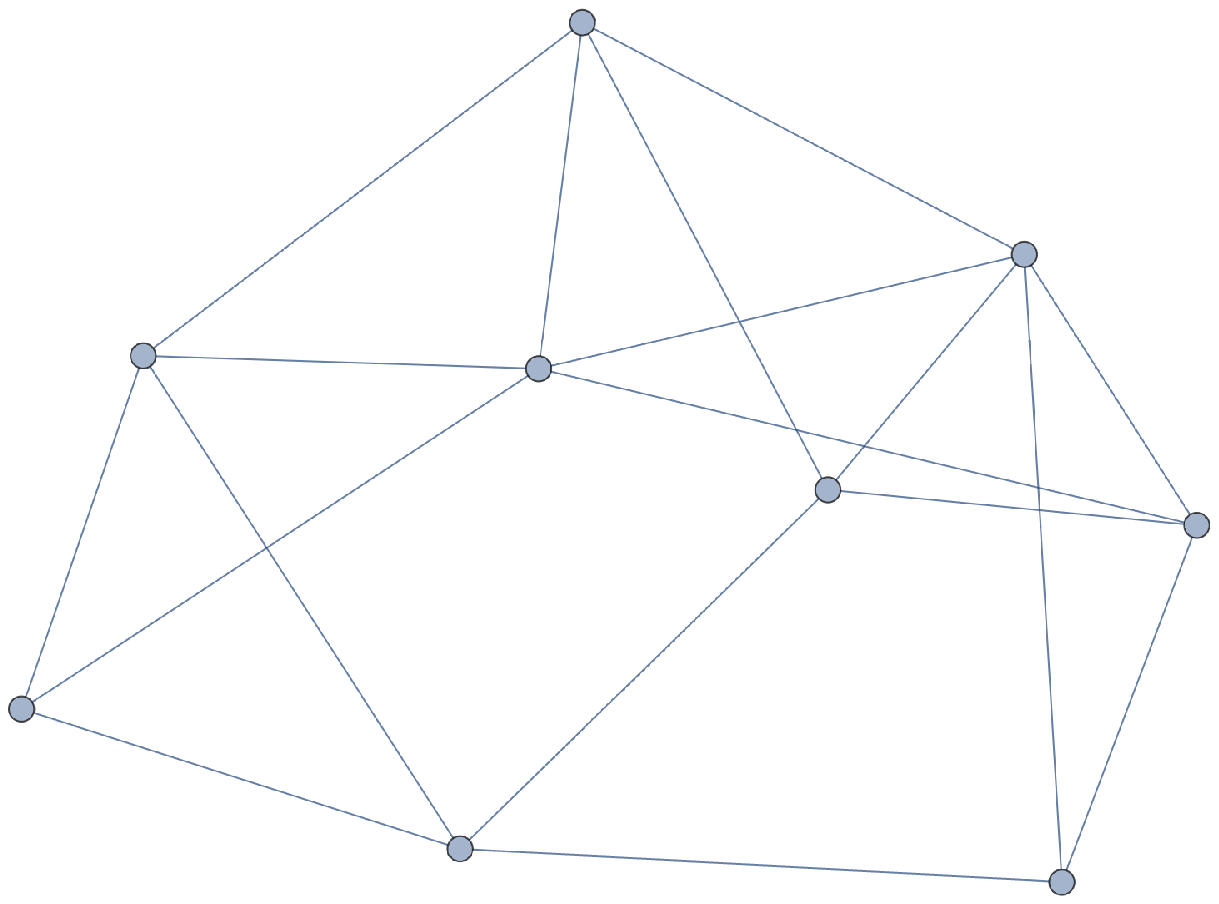}
   \subfigure[]{}
   \includegraphics[scale=\SCLL]{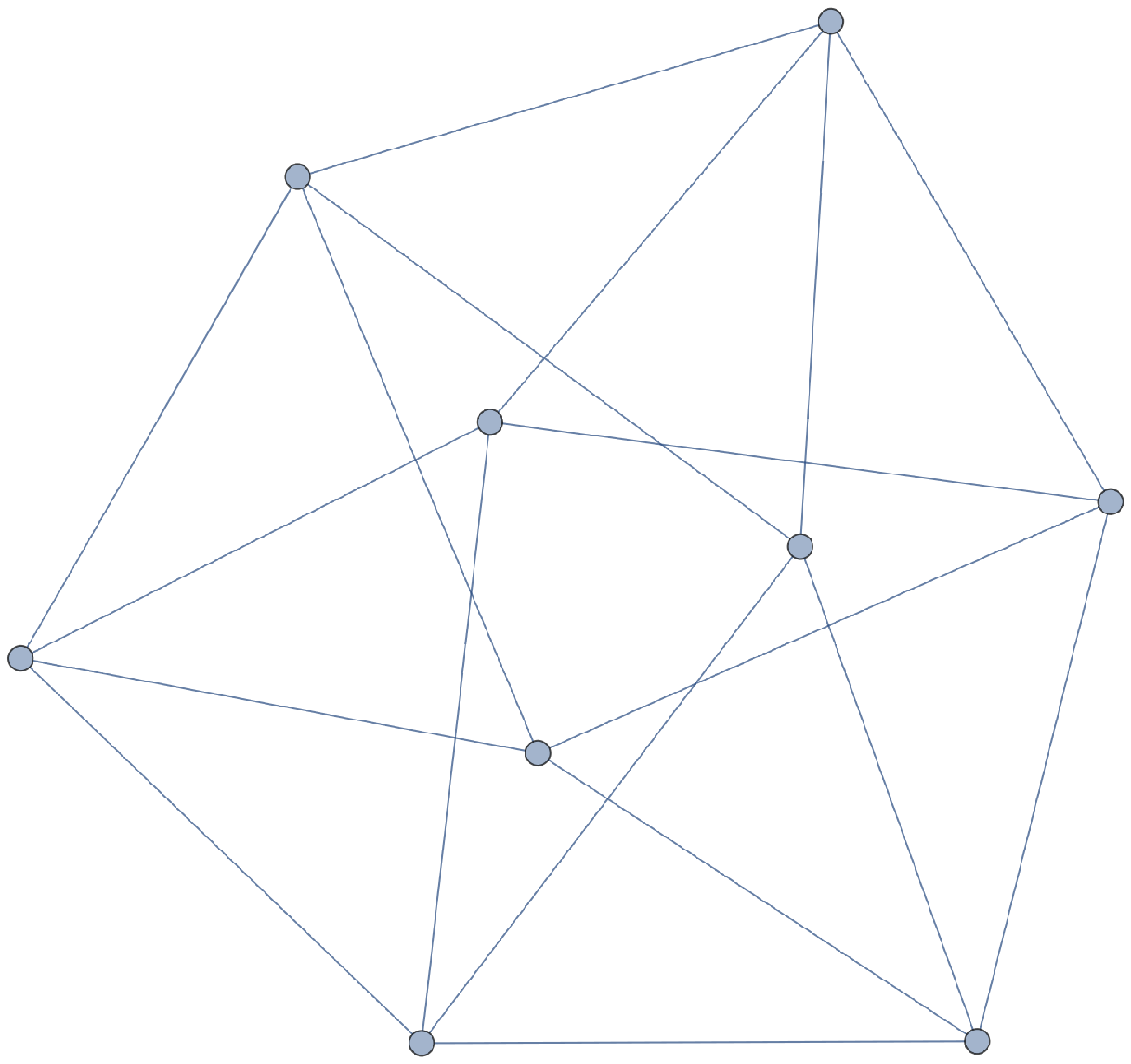}

   \caption{
       Four different graphs used in calculating the  critical parameters of the TQPT of $\mathbb{Z}_2$ LGT. (a) $G_1: N_v=9, N_e=18, N_{hc} =0$. (b) $G_2: N_v=9, N_e=18, N_{hc} =10$. (c) $G_3: N_v=9, N_e=18, N_{hc} =30$. (d) $G_4$: the graph corresponding to the $3\times 3$ torus lattice, $N_v=9, N_e=18, N_{hc} =48$.
   }
   \label{fig_hpc_list_graph}
\end{figure}

\begin{figure}[htb]
   \centering
   \subfigure[]{}
   \includegraphics[scale=\SCL]{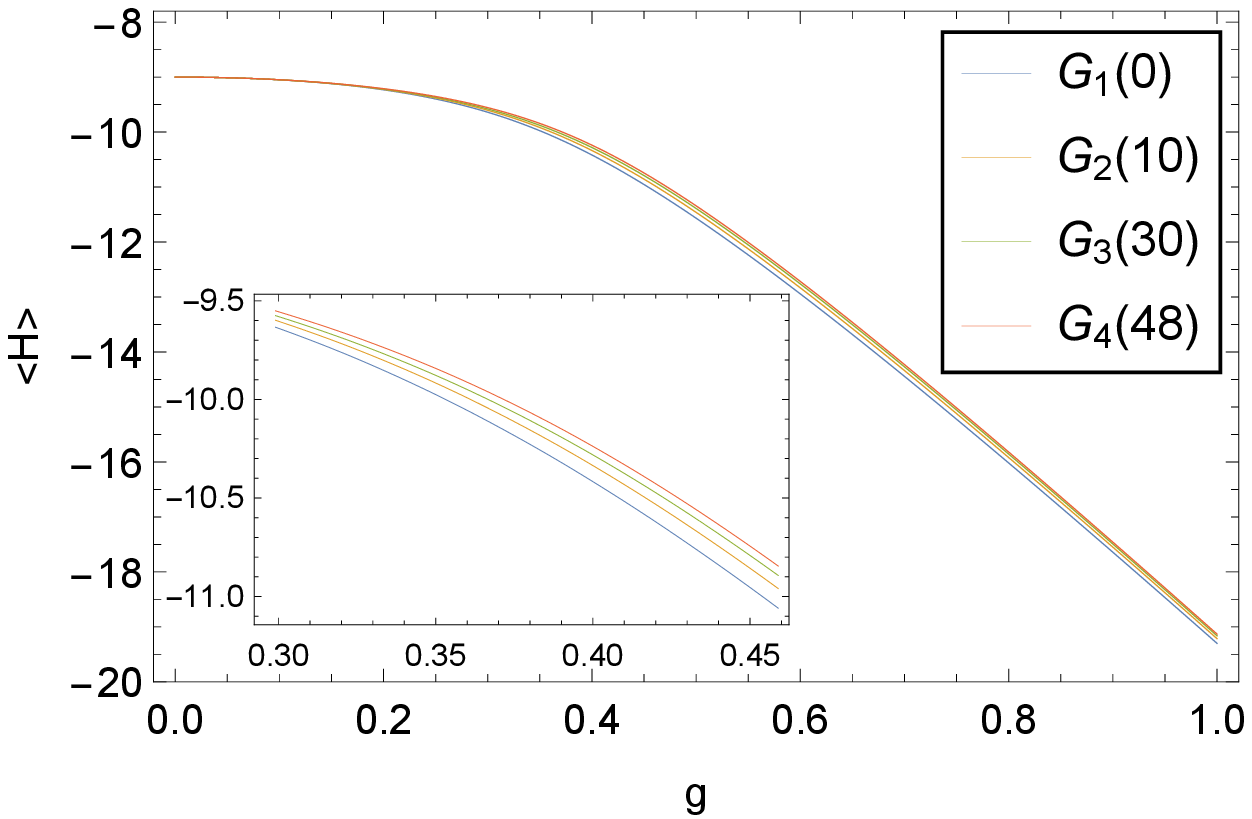}
   \subfigure[]{}
   \includegraphics[scale=\SCL]{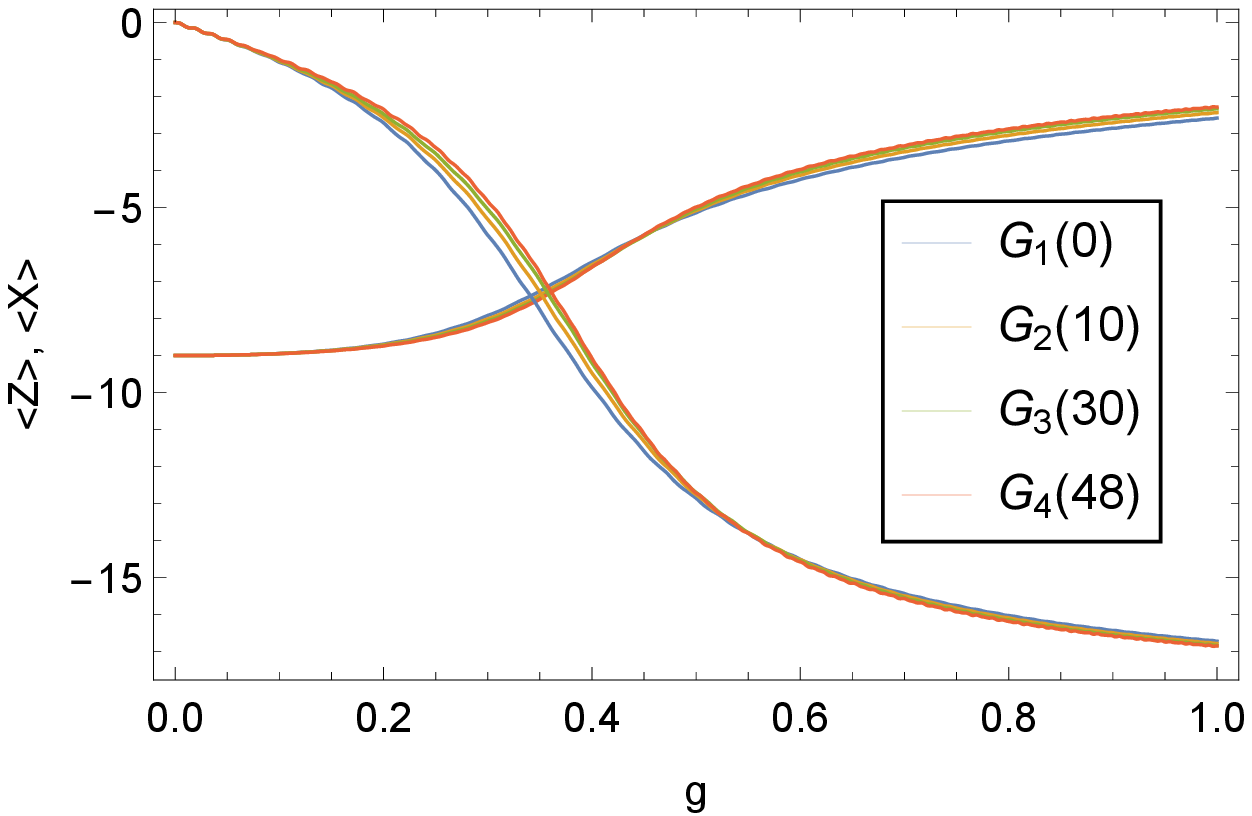}

   \subfigure[]{}
   \includegraphics[scale=\SCL]{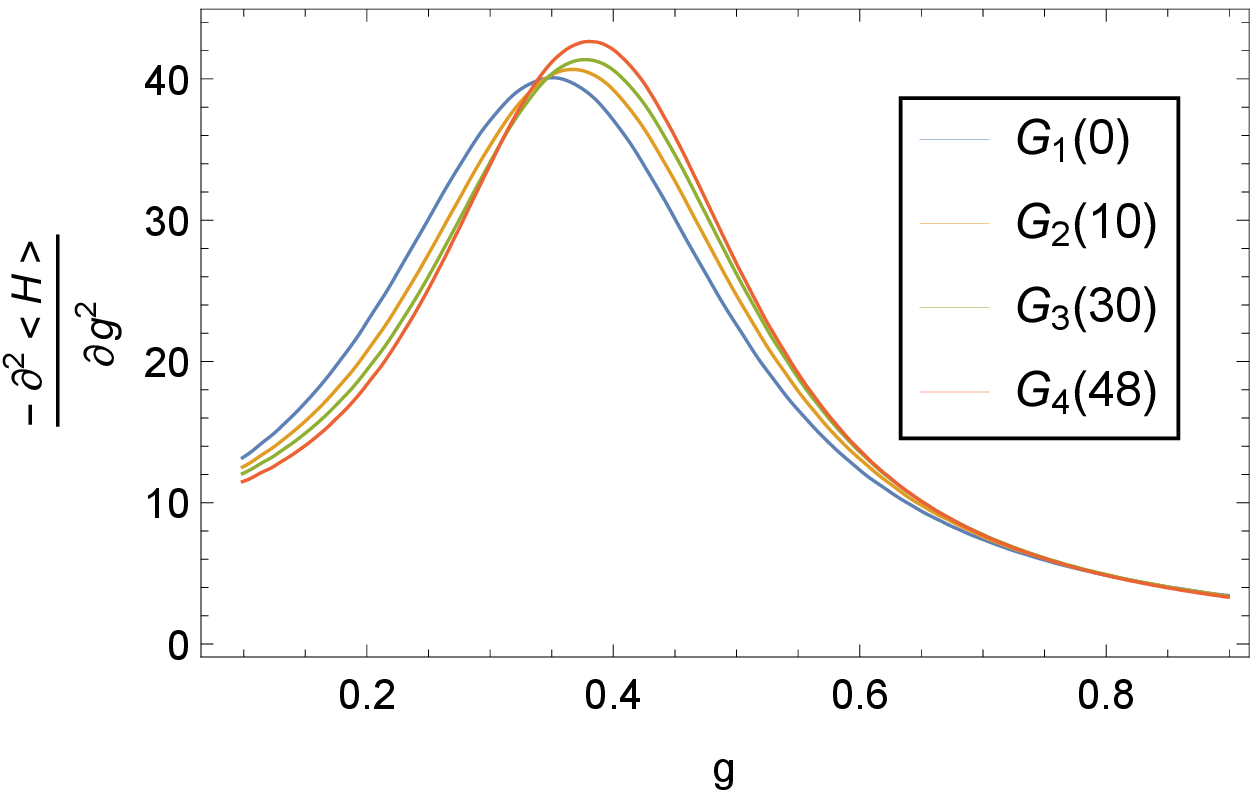}
   \subfigure[]{}
   \includegraphics[scale=\SCL]{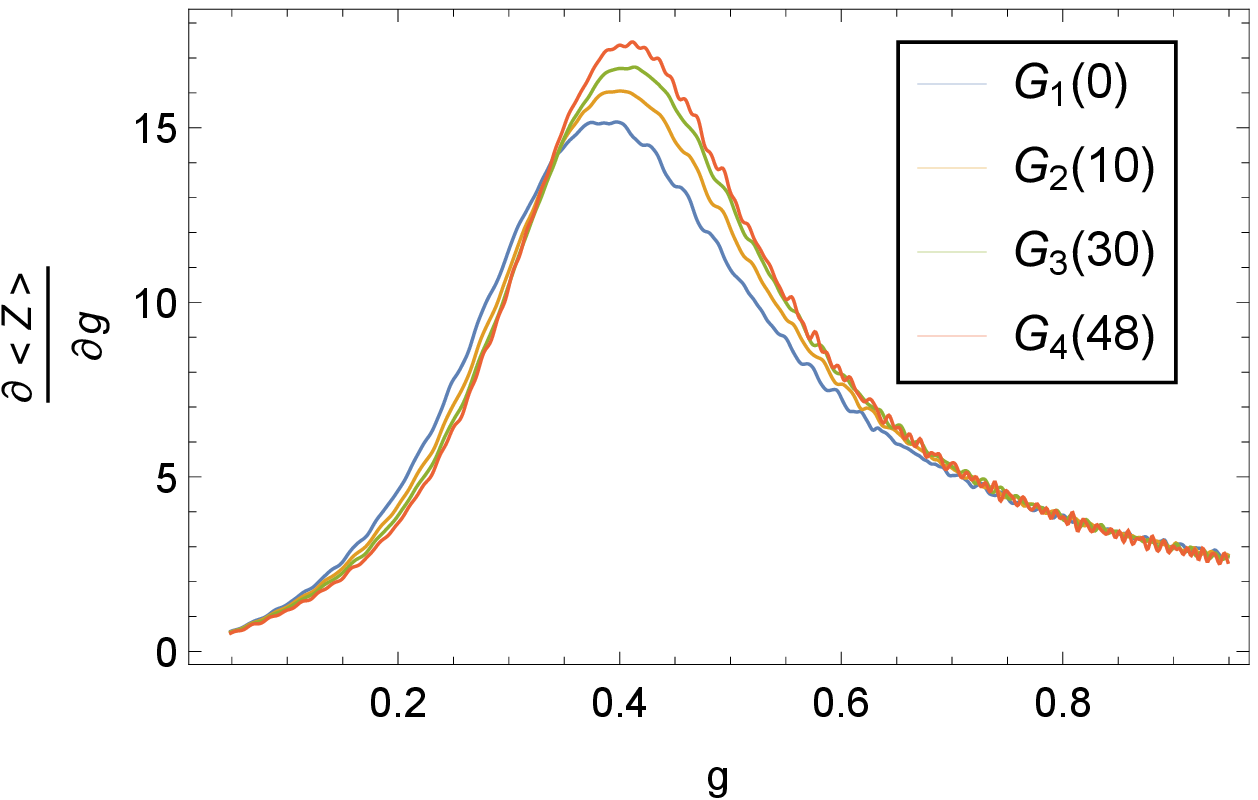}

   \caption{(a) $\langle H\rangle$  for four graphs denoted in terms of $N_{hc}$. (b) $\langle Z\rangle$  and $\langle X\rangle$ for four graphs denoted in terms of $N_{hc}$.  (c) The second derivatives  of $\langle H\rangle$ for the four  graphs, which can be used to determine the values of   $g_c^H$. (d) The first derivative  of $\langle Z\rangle$ for the four graphs, which can be used to determine the values of  $g_c^Z$.
   }
   \label{fig_hpc_list}
\end{figure}

Then we adiabatically  increase $g$ from 0 to 1, in steps with $g_s=0.001, t_s=0.1, n=100$, where $g_s$  is the variation of $g$,  $t_s$ and  $n$ are the time and the number of the symmetric Trotter substeps within each step, respectively. The total cumulative error is  $\varepsilon_{all} = \sum_{g=g_s}^1{\varepsilon_s(t,n,g)}$, which is less than $0.135\%$ for the graphs studied here.
We study   quantum phase transitions   on the  lattices corresponding to these  graphs. First,  $\braket{H}$,  $\braket{Z}$ and $\braket{Z}$  are obtained  as functions of  $g$.  Then we define  two critical parameters, which are easily determined, to represent the critical parameter $g_c$. One is the extremal point  $g_c^H$  of the second derivative of $\braket{H}$,  the other is   the extremal point $g_c^Z$ of the first derivative of  $\braket{Z}$.    $\lambda_c^H=\frac{1}{g_c^H}$, $\lambda_c^Z=\frac{1}{g_c^Z}$. As can be observed in Fig.~\ref{fig_hpc_list}(c-d),  $g_c^H$ and  $g_c^Z$ both increase with $N_{hc}$. This verifies  that the properties of the graph  directly affect  the characteristics of the quantum phase transition.

\begin{figure}[htb]
   \centering
   \subfigure[]{}
   \includegraphics[scale=\SCL]{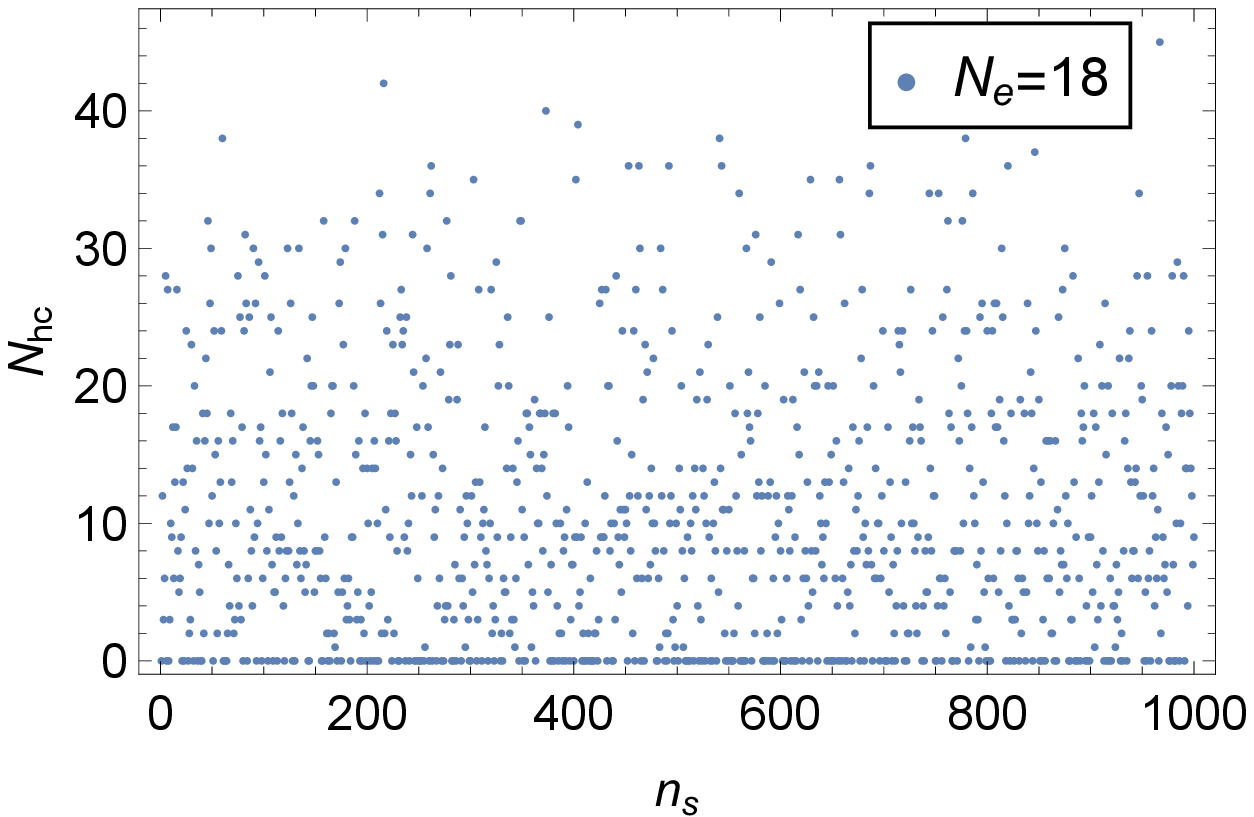}
   \subfigure[]{}
   \includegraphics[scale=\SCL]{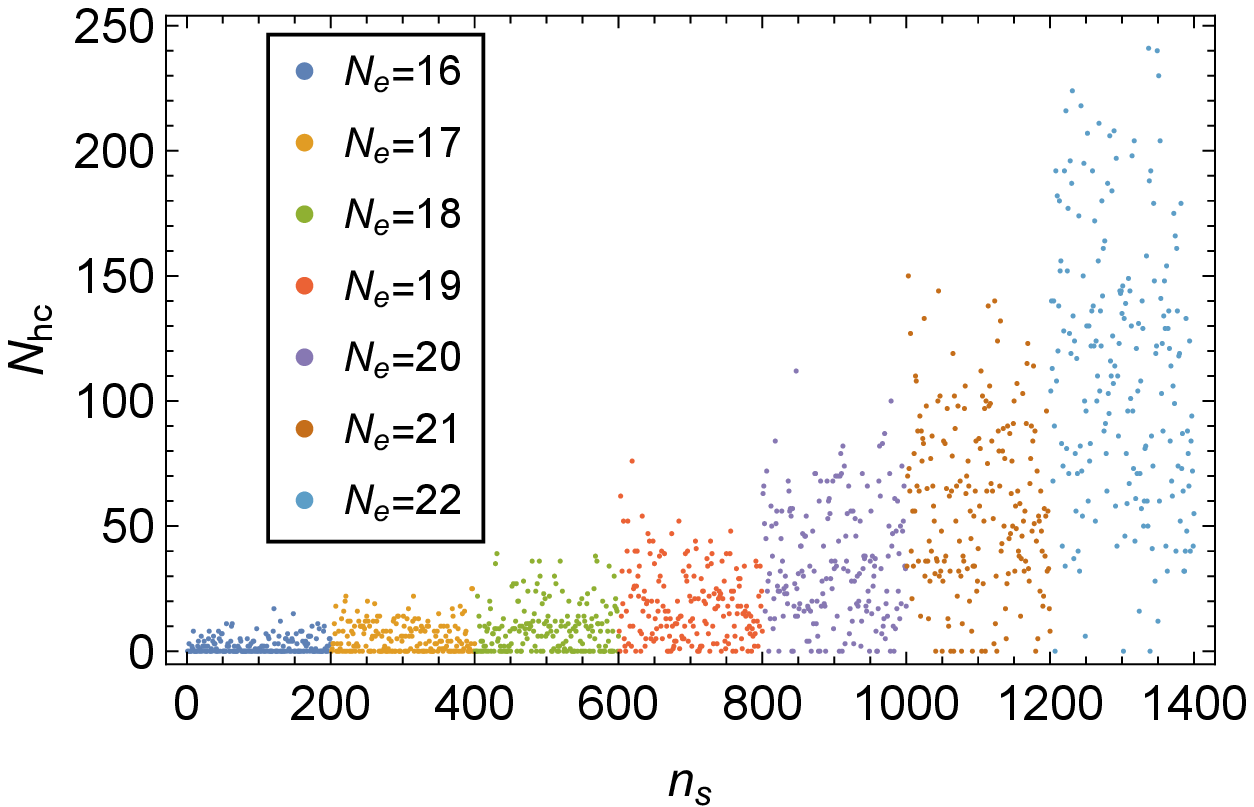}

   \caption{
      Distribution of the $N_{hc}$ among  samples of graphs. $n_s$ represents the number of the samples. (a) The case in which $N_e = 18$ in each sample. There are 1000 samples in total. (b) The case in which   $N_e$ varying from $16$ to $22$. There are 200 samples with each value of $N_e$, and 1400 samples in total.
   }
   \label{fig_hpc_sample}
\end{figure}

In order to study how $g_c^H$ and  $g_c^Z$ depend on  $N_{hc}$, we prepare two groups of undirected and unweighted connected graphs with $N_v=9$. The first group consists of 1000 samples, each with  $N_e=18$. We make the classical GPU demonstration of the adiabatic quantum simulation, and obtained  $N_{hc}$ for each graph.   The distribution of  $N_{hc}$  is shown in Fig.~\ref{fig_hpc_sample}(a). The second group consists of   $200$ graphs for  each value of  $N_e$ varying from $16$ to $22$. The distribution of  $N_{hc}$  is shown in the Fig.~\ref{fig_hpc_sample}(b).

\begin{figure}[htb]
   \centering
   \subfigure[]{}
   \includegraphics[scale=\SCL]{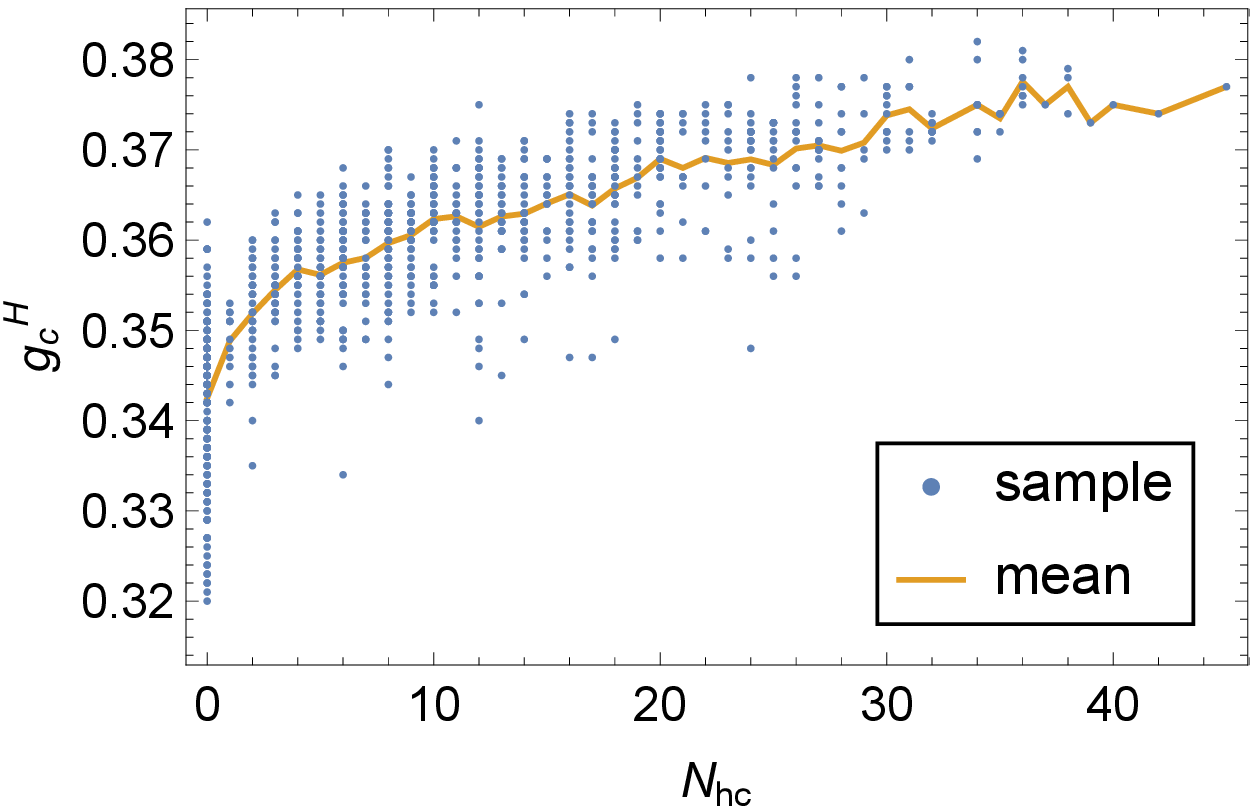}
   \subfigure[]{}
   \includegraphics[scale=\SCL]{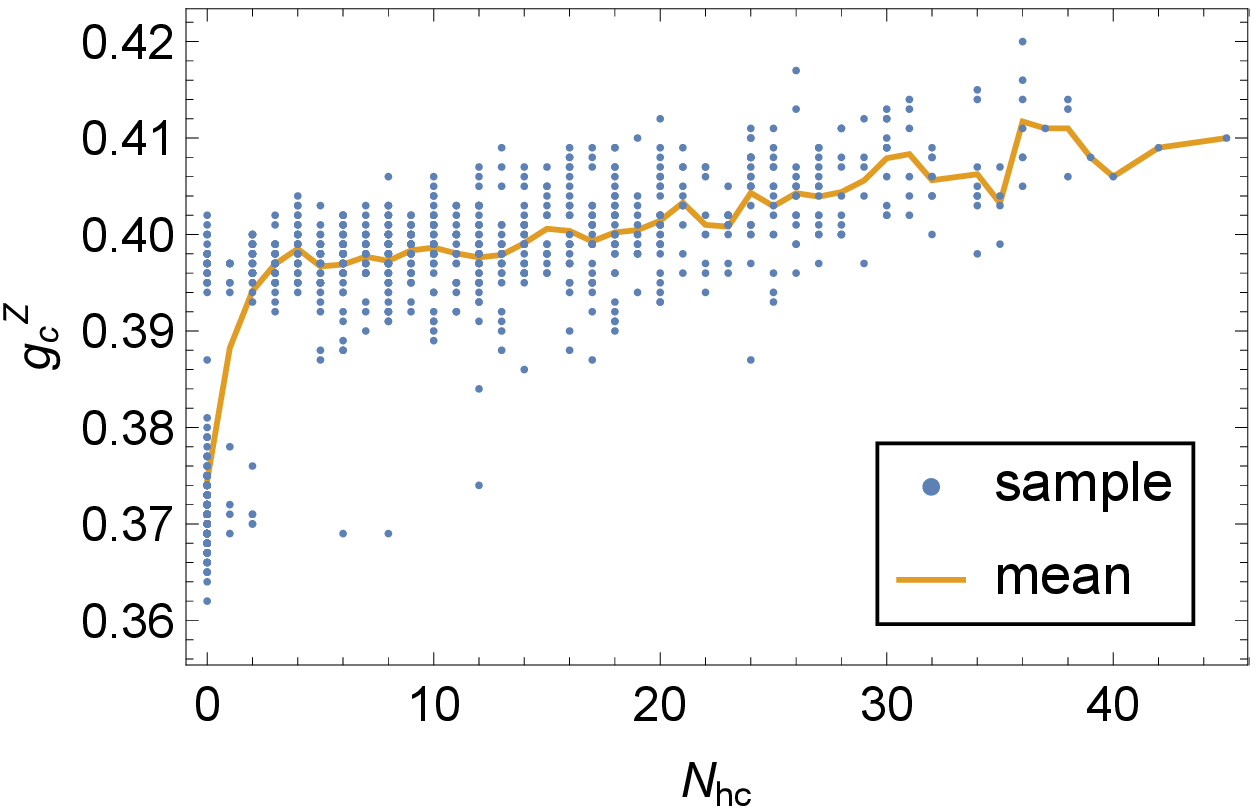}

   \subfigure[]{}
   \includegraphics[scale=\SCL]{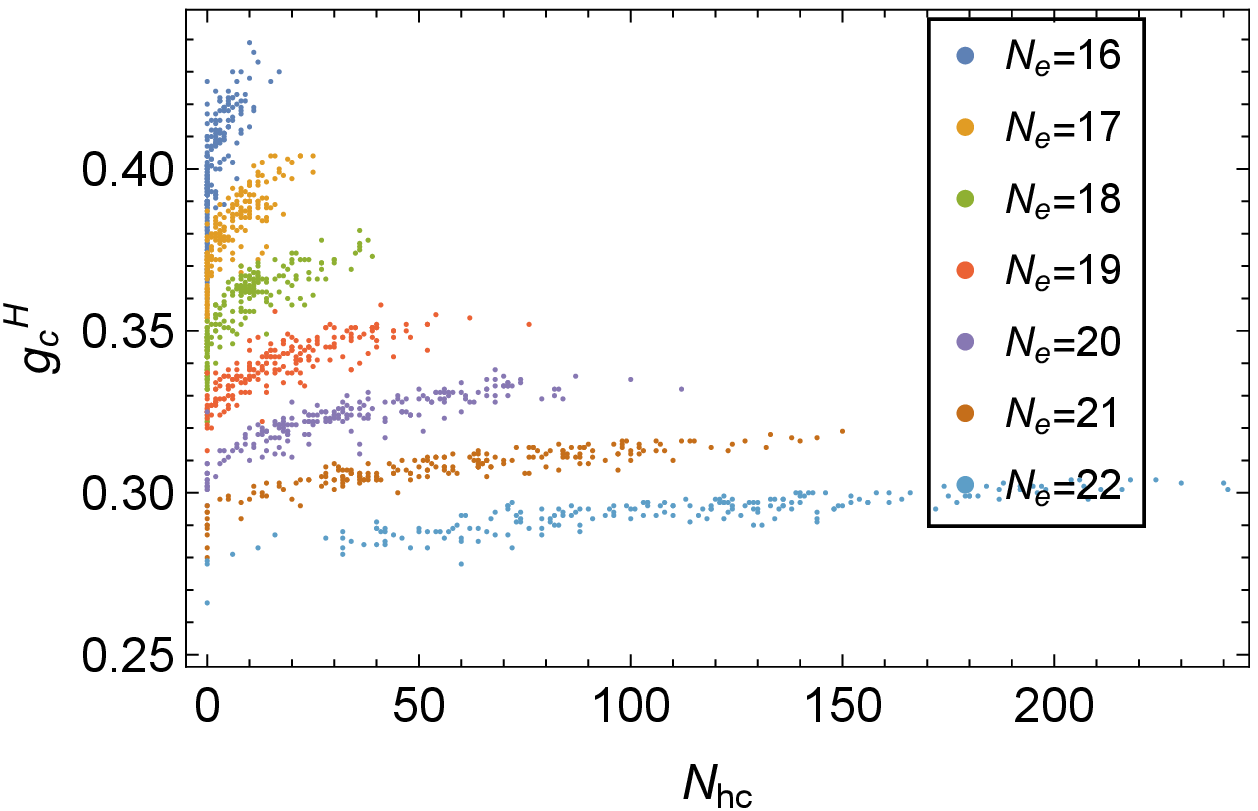}
   \subfigure[]{}
   \includegraphics[scale=\SCL]{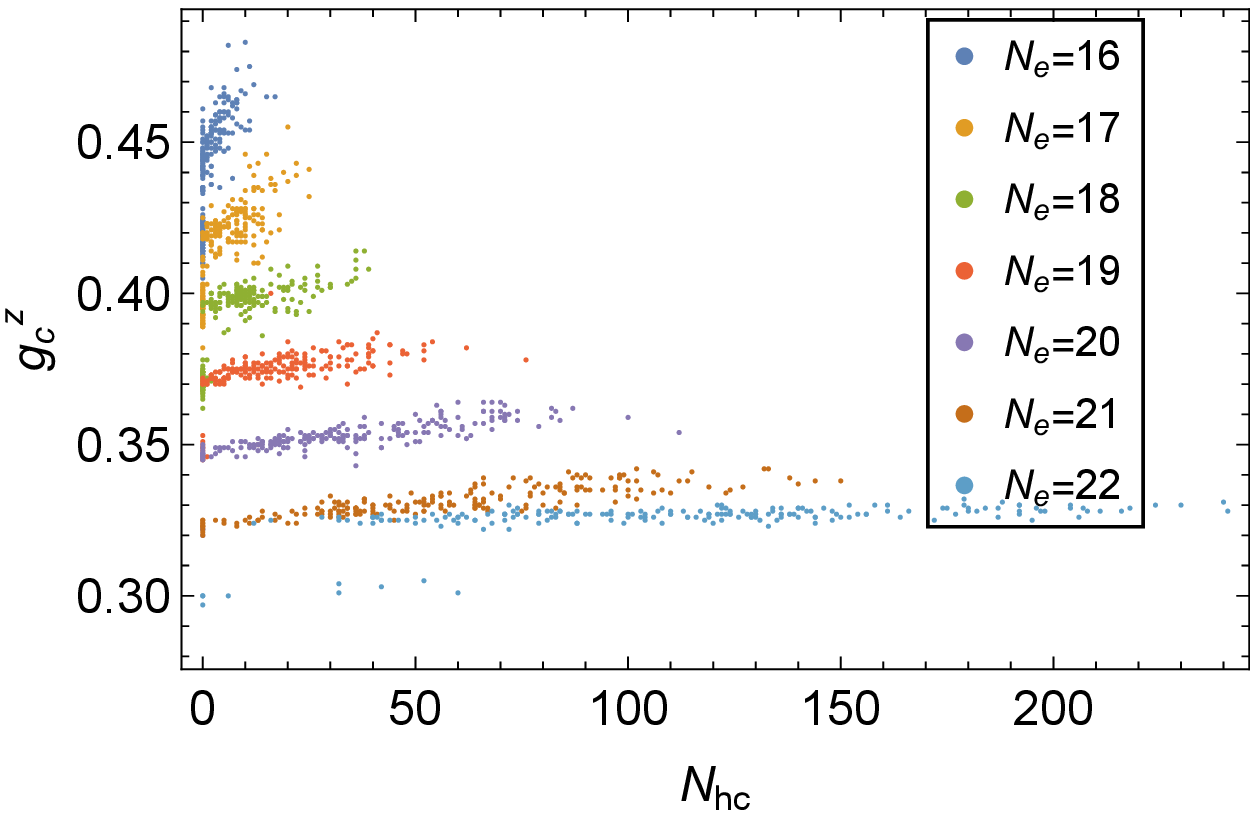}
   \caption{
      Dependence of the  critical parameters on  $N_{hc}$. (a) Dependence of  $g_c^H$  on $N_{hc}$ in graphs with  $N_e=18$. (b) Dependence of  $g_c^Z$  on $N_{hc}$ in graphs with  $N_e=18$.   (c) Dependence of  $g_c^H$  on $N_{hc}$ in graphs with  $N_e$ varying from    $16$ to $22$. (d) Dependence of  $g_c^Z$  on $N_{hc}$ in graphs with  $N_e$ varying from    $16$ to $22$.        }
   \label{fig_hpc_N_hc}
\end{figure}

It can be seen from  Fig.~\ref{fig_hpc_N_hc} that the average value of $g_c^H$ and $g_c^Z$ increase  steadily with $N_{hc}$ when $N_e$ is fixed. Especially, when $N_{hc}$ = 0,  $g_c^H$ and $g_c^Z$ are very small. This shows that $N_{hc}$ has a significant effect on $g_c$. This effect can help determine $N_{hc}$ without searching the HCs of the graph. The time complexity required to reach TQPT is $O_1$. $g_c^H$ and $g_c^Z$ may   decrease linearly with $N_e$,  as discussed below. Together with $N_v \leq N_e \leq N_v(N_v-1)/2$, these properties imply  that HC problem in these small graphs can be solved in polynomial time.

\begin{figure}[htb]
   \centering
   \subfigure[]{}
   \includegraphics[scale=\SCL]{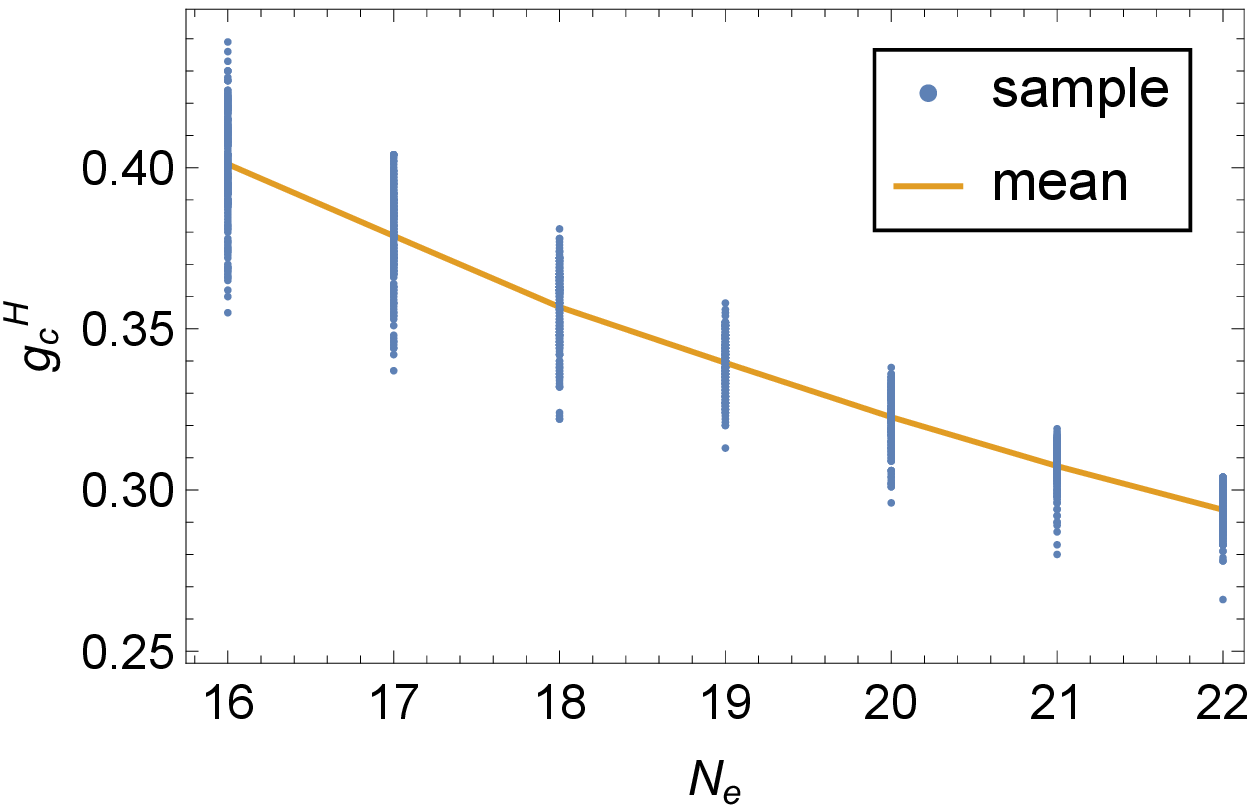}
   \subfigure[]{}
   \includegraphics[scale=\SCL]{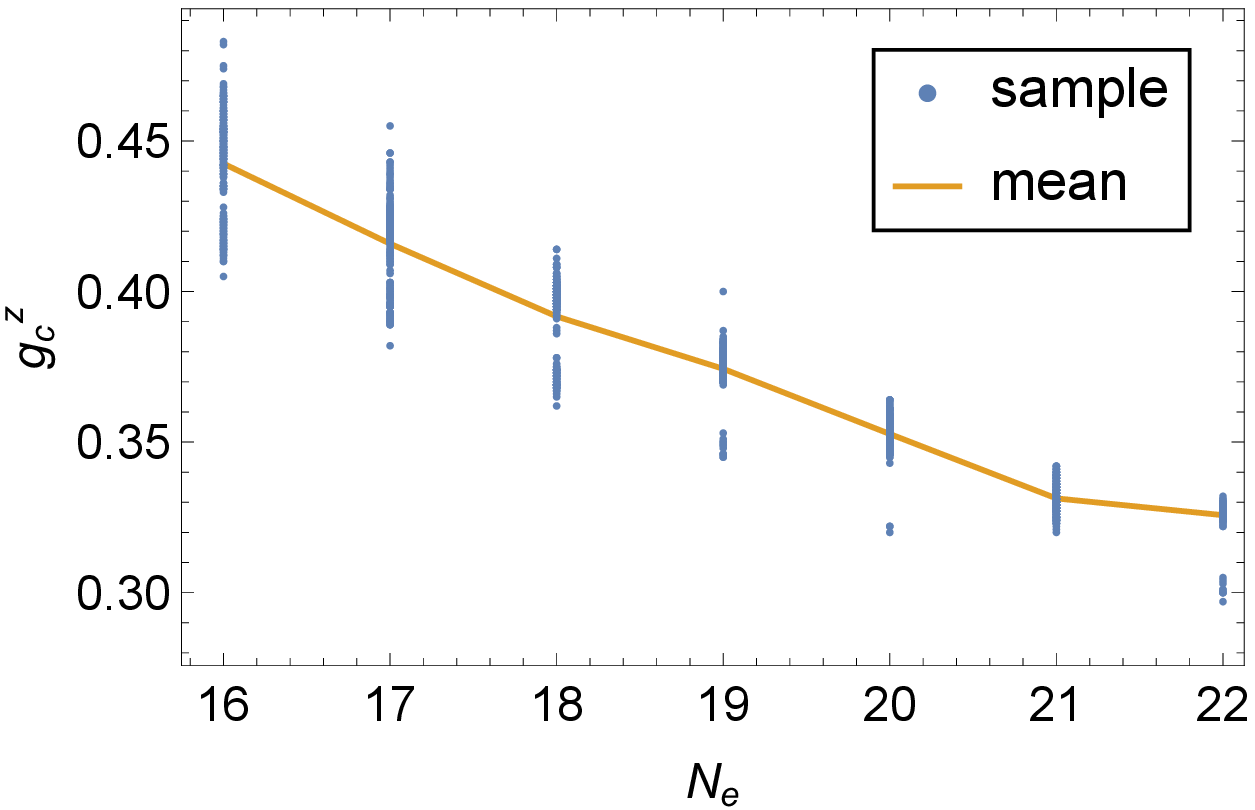}

   \subfigure[]{}
   \includegraphics[scale=\SCL]{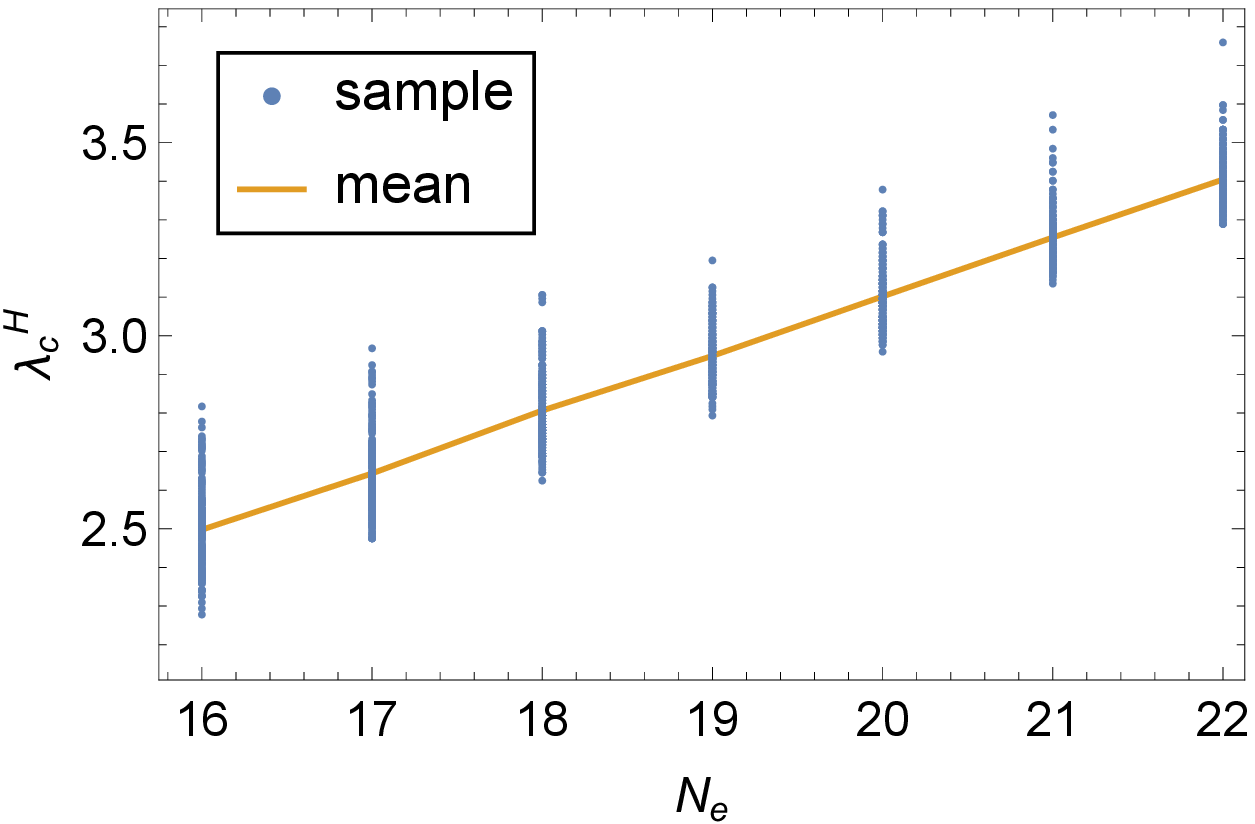}
   \subfigure[]{}
   \includegraphics[scale=\SCL]{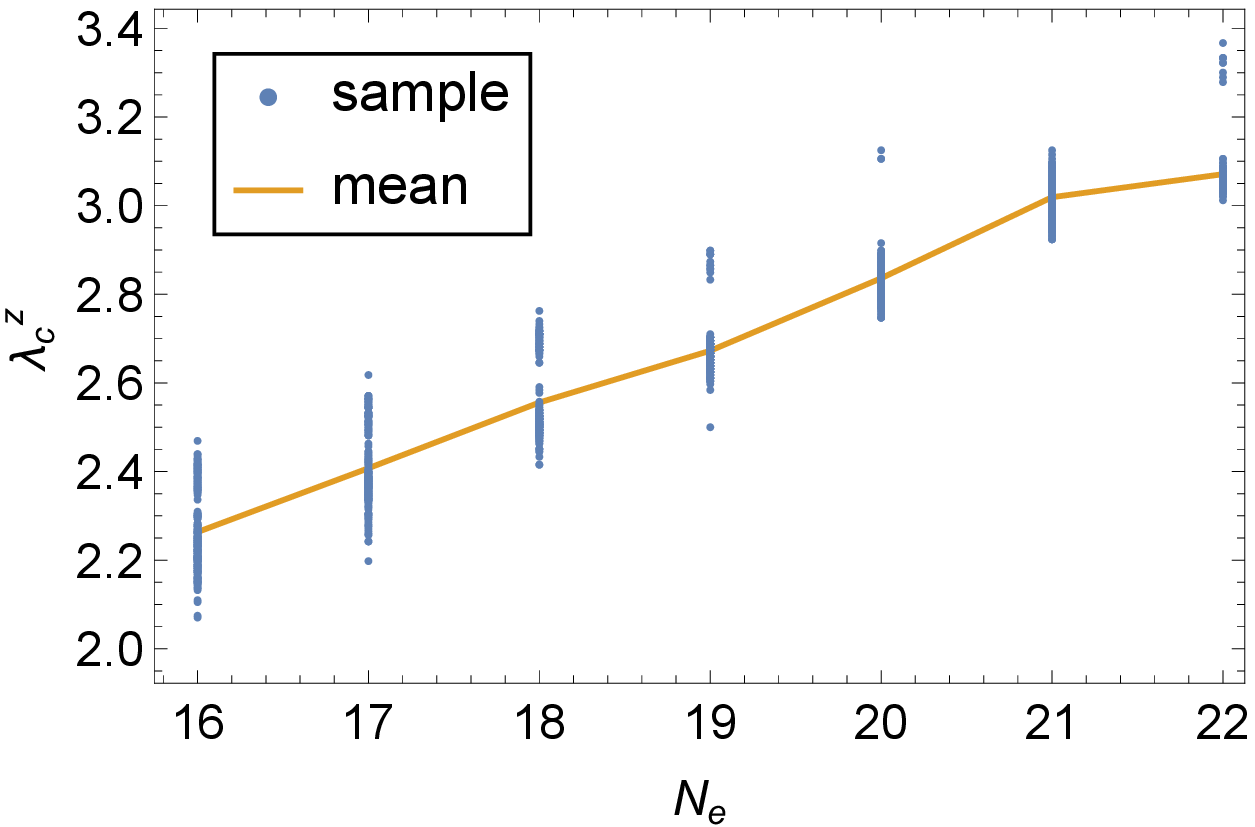}

   \caption{
     The dependence of the critical parameters  on   $N_{e}$.      (a)  The dependence of $g_c^H$ on $N_{e}$ in the  samples with $N_e$ varying from $16$ to $22$.  (b)  The dependence of $g_c^Z$ on $N_{e}$ in the  samples with $N_e$ varying from $16$ to $22$. (c)  The dependence of $\lambda_c^H$ on $N_{e}$ in the  samples with $N_e$ varying from $16$ to $22$.  (d)  The dependence of $\lambda_c^Z$ on $N_{e}$ in the  samples with $N_e$ varying from $16$ to $22$.
   }
   \label{fig_hpc_all_N_edge}
\end{figure}

We have also studied   the effect of $N_{e}$ on the critical parameters, by using the second group of graphs with the same $N_v$ value but different $N_e$ values. As can be seen in the Fig.~\ref{fig_hpc_all_N_edge}, the average values of   $g_c^H$  and  $g_c^Z$  decrease steadily with $N_e$ when $N_v$ is fixed. The larger the $N_e$ value, the smaller the critical parameters. In other words, the average value of $\lambda_c = \frac{1}{g_c}$ increase linearly with $N_e$ when $N_v$ is fixed. Thus $O_1$ increases with $N_e$ polynomially. On average, when $N_v$ is fixed, the connectivity between vertices of a graph is proportional to $N_e$.


\begin{figure}[htb]
   \centering
   \subfigure[]{}
   \includegraphics[scale=\SCL]{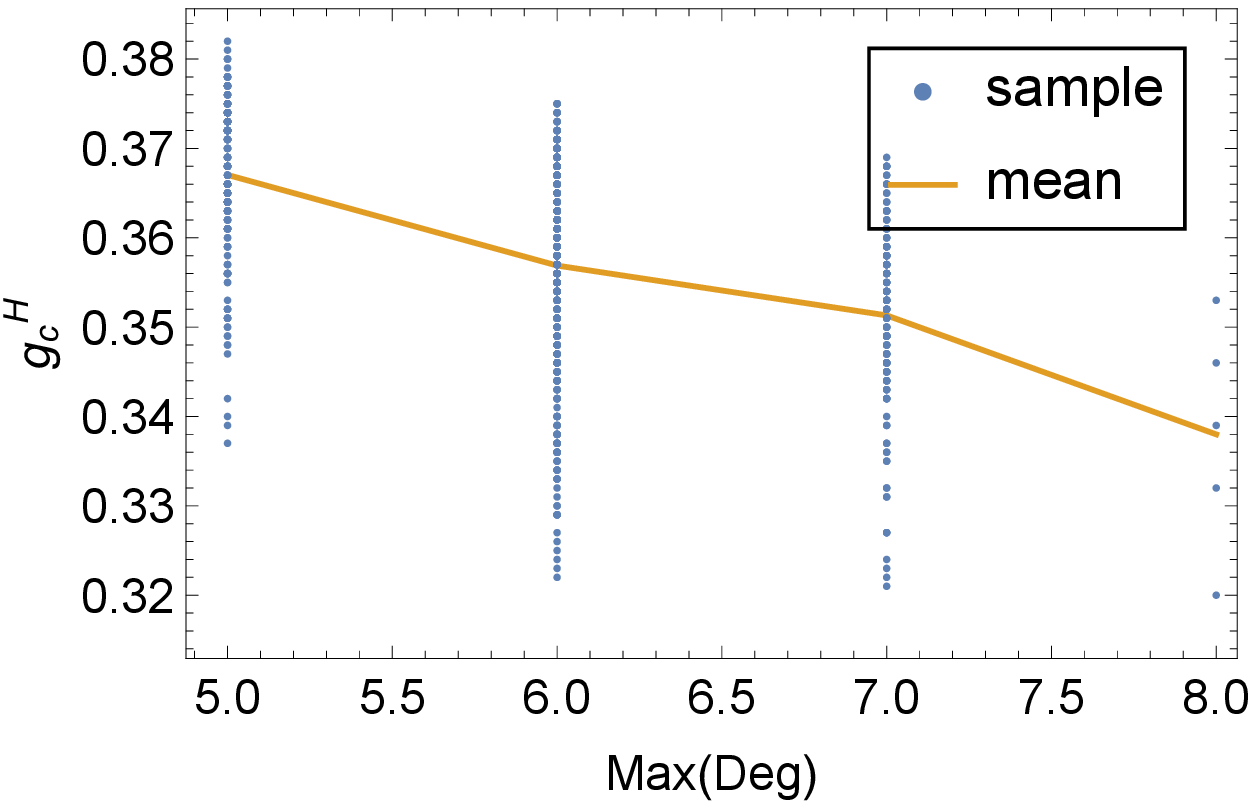}
   \subfigure[]{}
   \includegraphics[scale=\SCL]{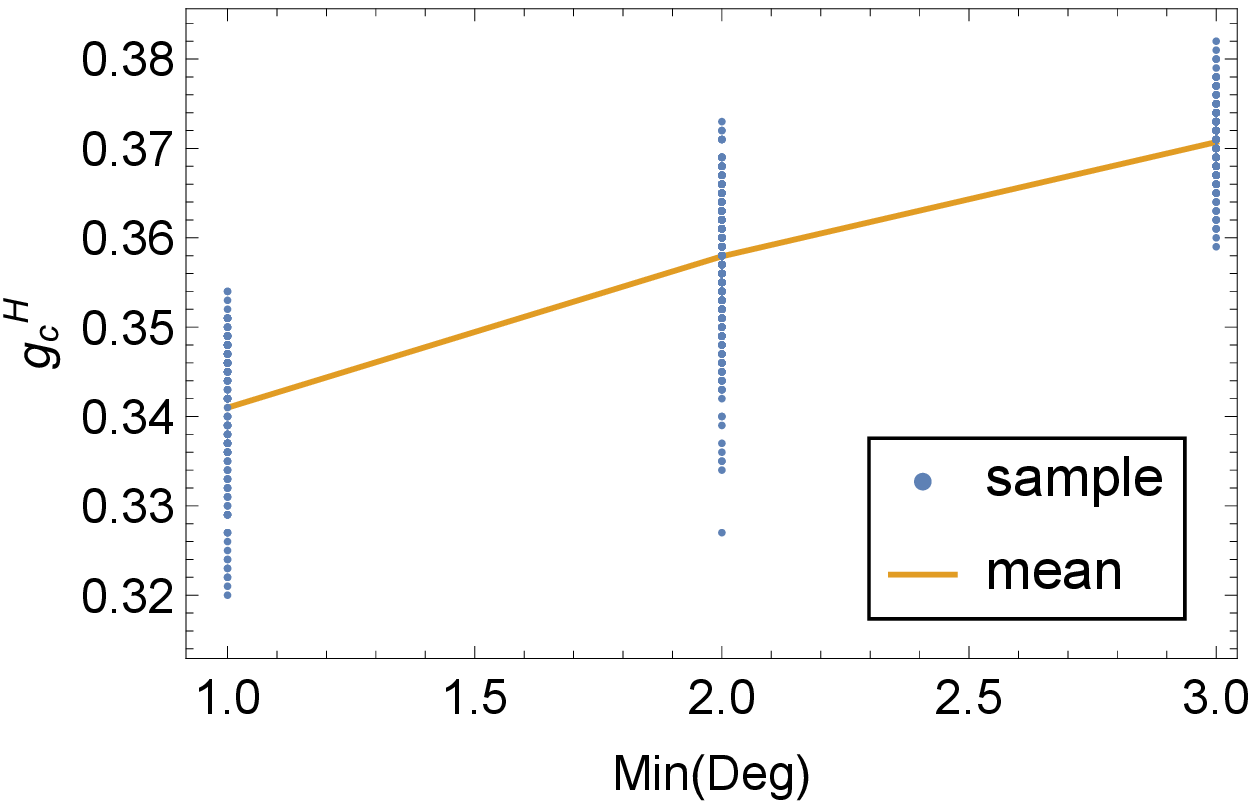}

   \subfigure[]{}
   \includegraphics[scale=\SCL]{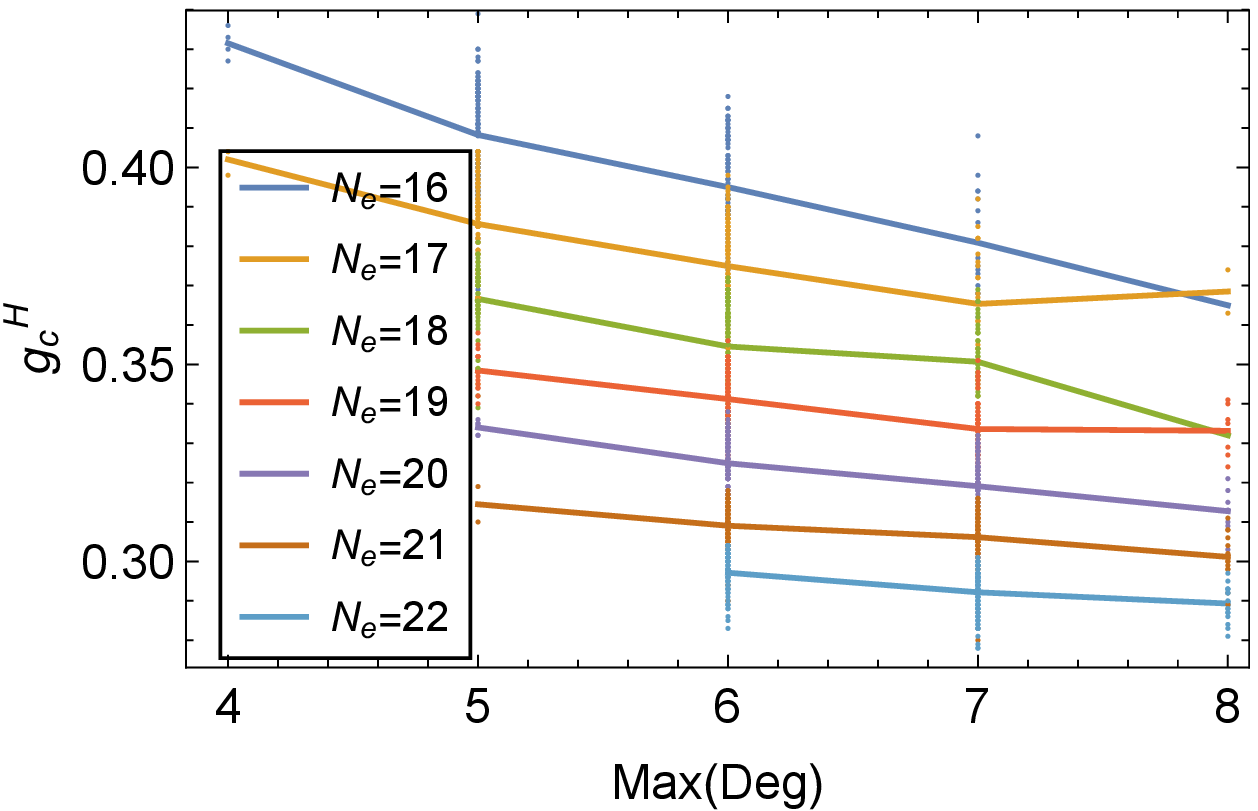}
   \subfigure[]{}
   \includegraphics[scale=\SCL]{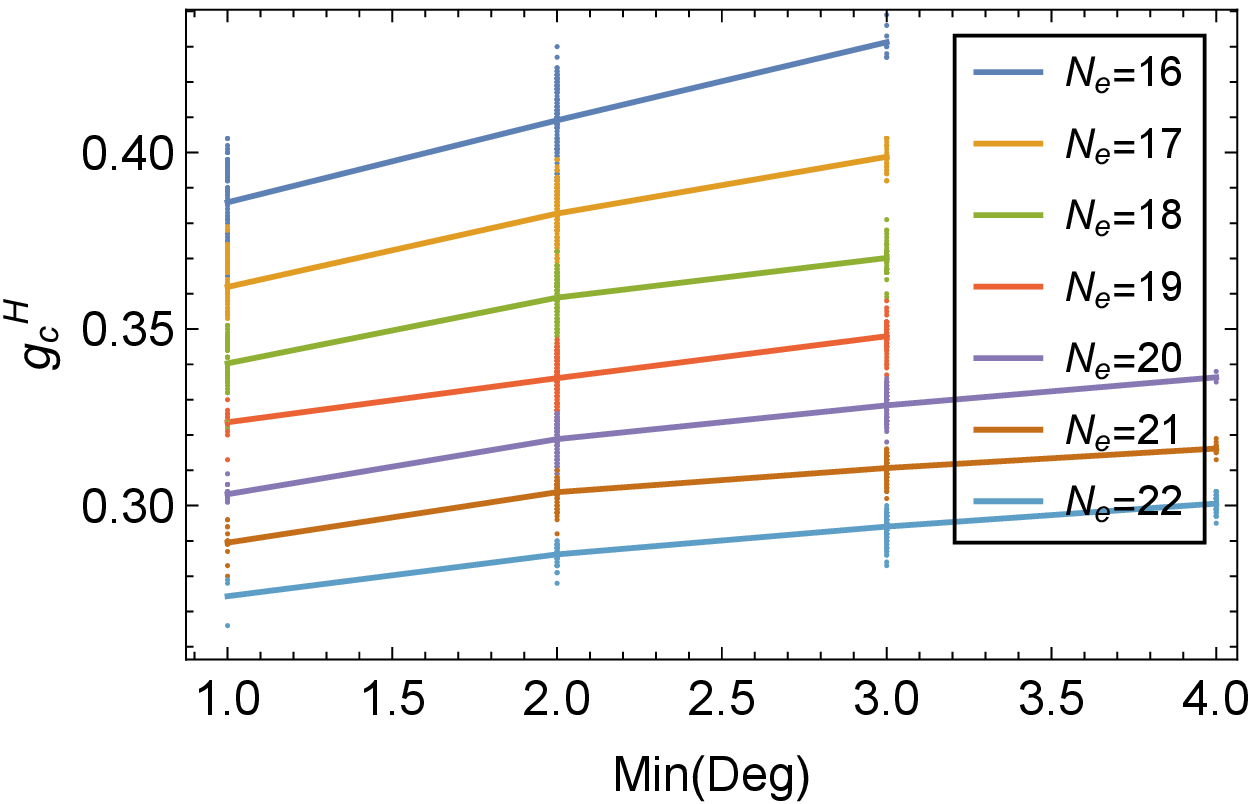}
   \caption{
     The dependence of the critical parameters  on  the degree of vertices.  (a)  The dependence of $g_c^H$ on  $Max(Deg)$   in the  samples with $N_e=18$. (b) The dependence of $g_c^H$ on $Min(Deg)$ in the  samples with $N_e=18$.    (c)  The dependence of $g_c^H$ on  $Max(Deg)$  in the  samples with $N_e$ varying from $16$ to $22$.  (d)  The dependence of $g_c^Z$ on $Min(Deg)$ in the  samples with $N_e$ varying from $16$ to $22$.
   }
   \label{fig_hpc_deg}
\end{figure}

As can be seen in  Fig. \ref{fig_hpc_deg},  when $N_e$ and $N_v$ are fixed, the average values of the critical parameters  decrease  with the maximal degree of the vertices $Max(Deg)$,  while increases with the  minimal degree of the vertices  $Min(Deg)$.
On average, when $N_v$ and $N_e$ are fixed, the larger $Max(Deg)$, the smaller $(Min(Deg))$. So the results in Fig.~\ref{fig_hpc_deg}(a,c) and Fig.~\ref{fig_hpc_deg}(b,d) are consistent.

\begin{figure}[htb]
   \centering
   \subfigure[]{}
   \includegraphics[width=0.45 \textwidth]{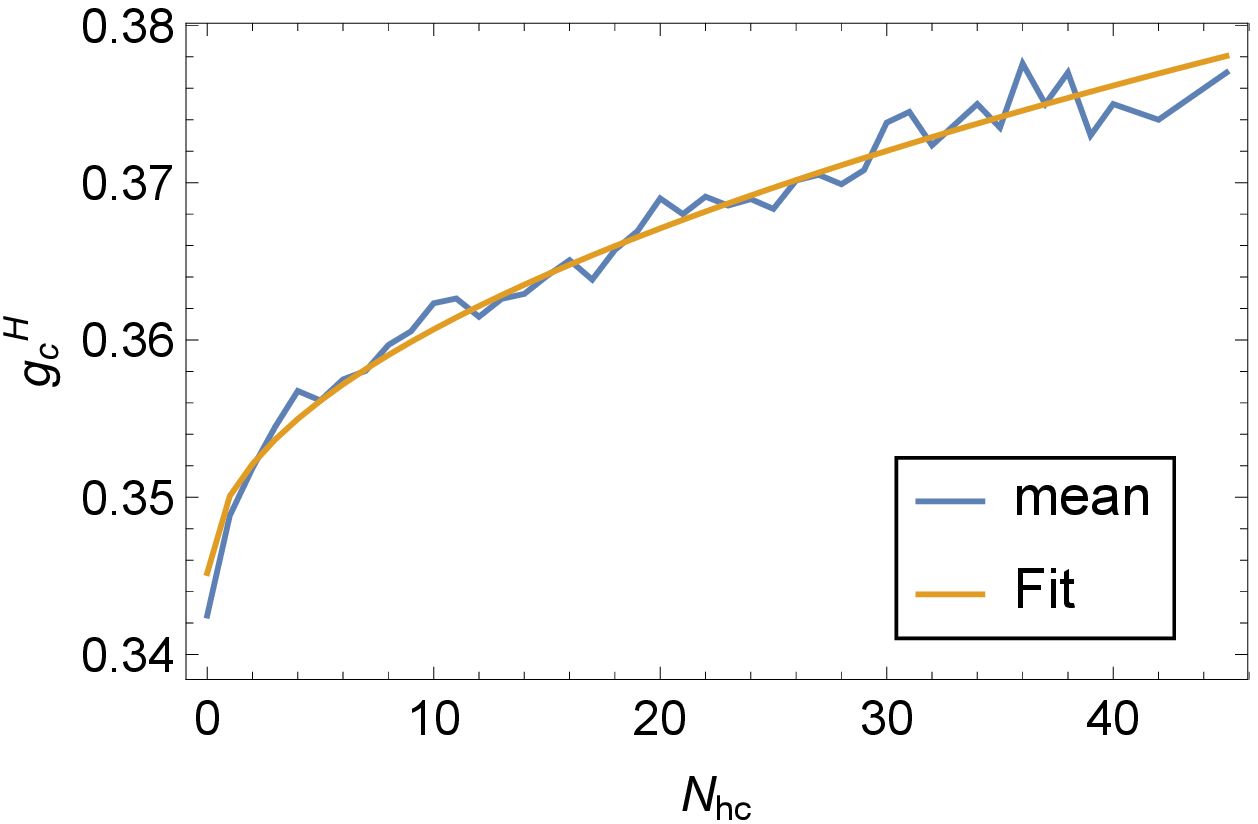}
   \subfigure[]{}
   \includegraphics[width=0.45 \textwidth]{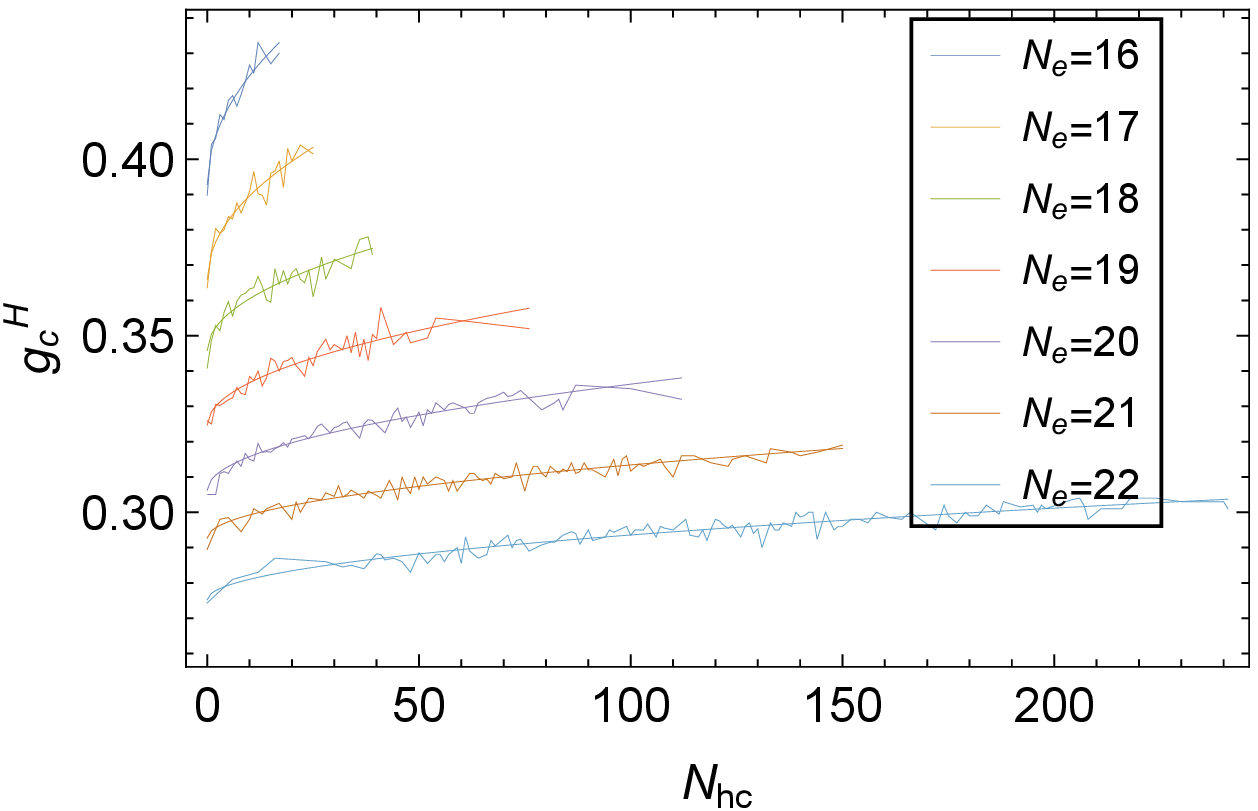}

   \subfigure[]{}
   \includegraphics[width=0.45 \textwidth]{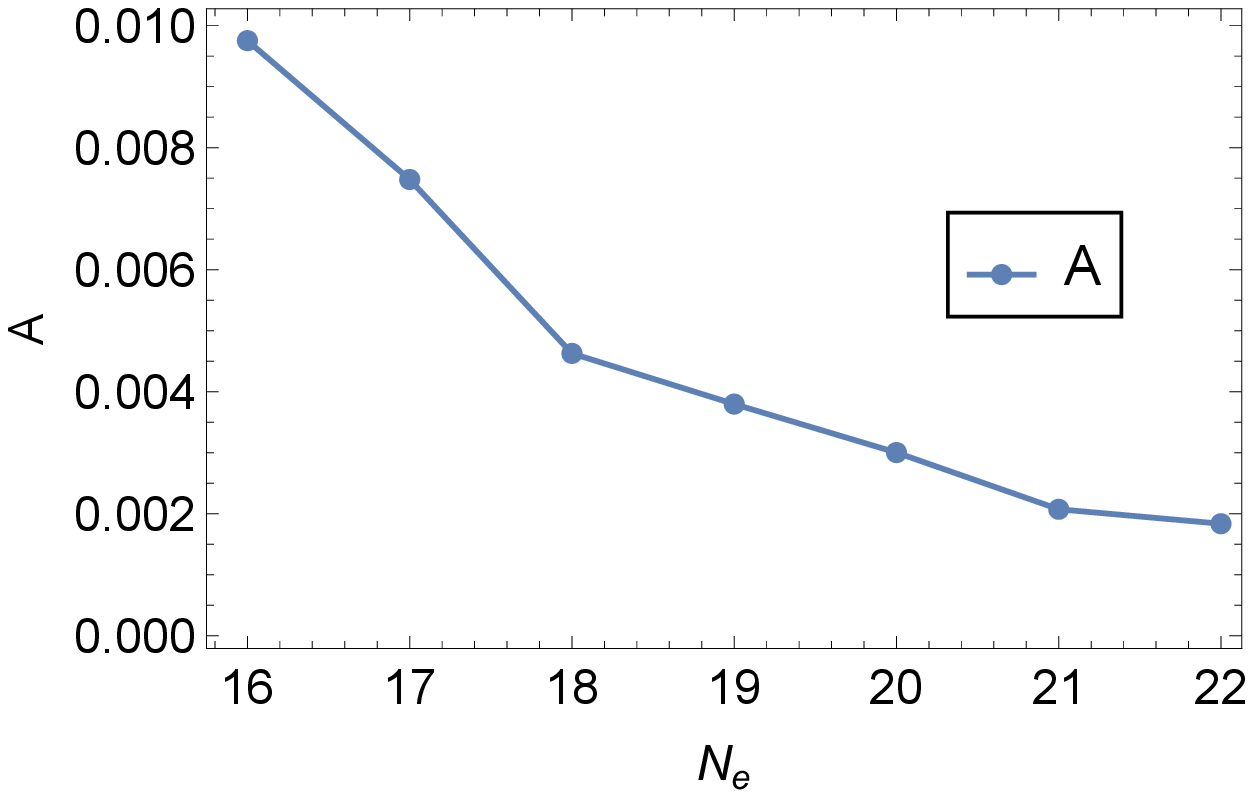}
   \subfigure[]{}
   \includegraphics[width=0.45 \textwidth]{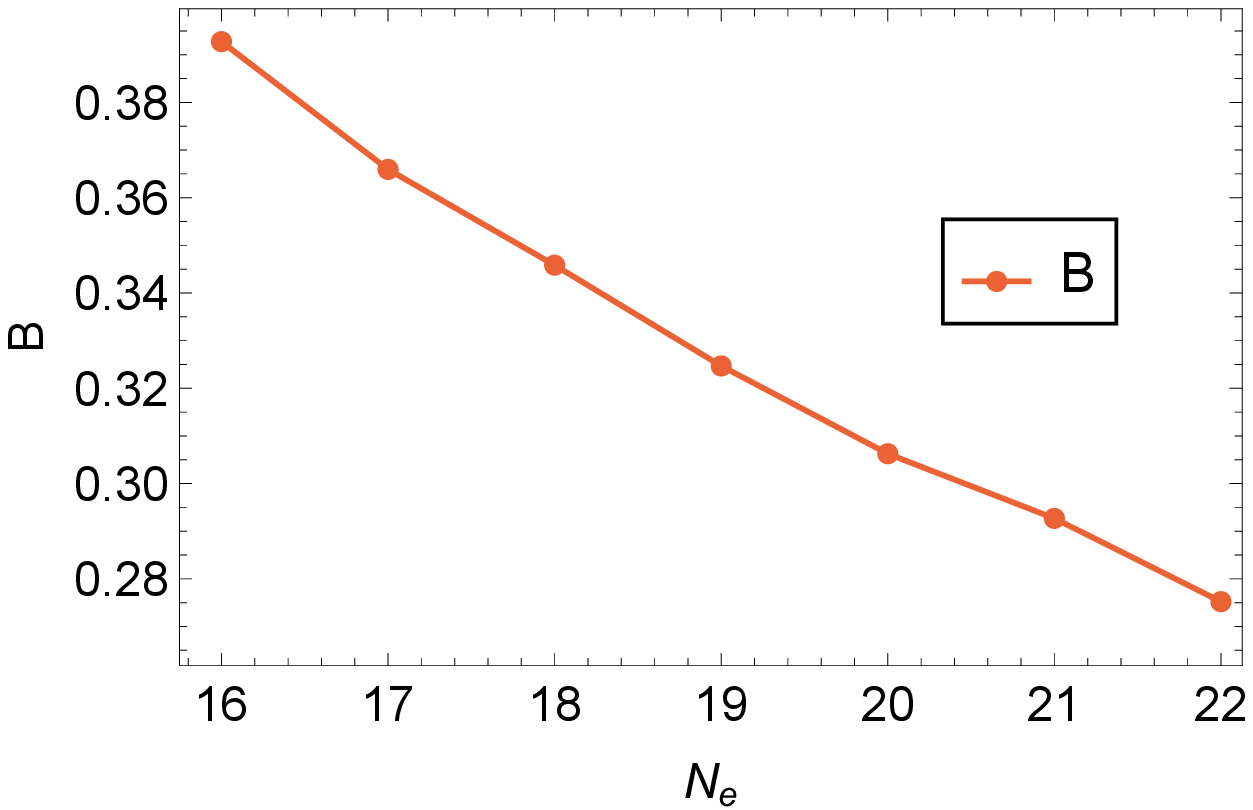}
   \caption{ The average value of  $g_c^H$ as a function of $N_{hc}$, fitted by $g_c^H = A\sqrt{N_ {hc}} + B $. (a) The  samples with $N_e=18$.   $A=0.0049005$, $B=0.345178$. (b) The sample with $N_e$ varying from $16$ to $22$, with $A$ and $B$ depending on $N_e$. (c) Dependence of $A$ in (b) on $N_e$. (d)Dependence of $B$ in (b) on $N_e$.}
   \label{fig_hpc_N_hc_fit}
\end{figure}

Furthermore, we find that  the relation between the average values of  $g_c^H$  and  $N_{hc}$ can be very well fitted quantitatively as
\begin{equation}
   g_c^H = A \sqrt{N_{hc}} + B, \label{gch}
\end{equation}
as shown  in Fig.~\ref{fig_hpc_N_hc_fit}   for the two groups of graph samples.
On the other hand, the  relation between $\lambda_c^H$ and $N_e$ can be very well fitted as
\begin{equation}
\lambda_c^H = 0.1513*N_e + 0.007536, \label{lch}
\end{equation}
as shown in Fig.\ref{fig_hpc_lam_fit}.
\eqref{gch} and  \eqref{lch} are consistent. Due to the limitation of computing power, $N_e$ is still relatively small in our simulated graphs, so the linear relationship needs verification in larger graphs.

In conclusion, we develop a novel approach to graph problem, by defining $\mathbb{Z}_2$ LGT on the lattice of which the graph is the dual lattice. Moreover,   we  present a   quantum adiabatic  algorithm with time complexity \COAO,   to  obtain the closed-string condensate  emerging at the  TQPT  of the  $\mathbb{Z}_2$ LGT,  and find that the HC number $N_{hc}$ of in the graph has a significant effect on the TQPT critical parameter $g_c$,  providing  a novel quantum algorithm for HC problem. Thereby  a  new approach to HC problem and P versus NP in quantum computing is proposed. Given the importance of graph theory in mathematics and computer science, our approach may also be useful to, say, deep learning, represented by deep neural networks and are being integrated with graphical  models~\cite{Johnson2016,HWang2016},  as well as  quantum deep learning~\cite{QBM_PRX,Lloyd2017}. This work also suggests a new direction of research connecting    graph problems with  topological quantum matter.

This work was supported by National Natural  Science Foundation of China (Grant No. 12075059).

\bibliography{qzt_SN}

\end{document}